\documentclass[aps,prd,letterpaper,twocolumn,superscriptaddress,showpacs,amsmath,amssymb,nofootinbib,floatfix]{revtex4-1}
\usepackage{graphicx}  
\graphicspath{ {Figures/} } 
\usepackage{bm}     
\usepackage{amssymb}   
\usepackage{amsmath}   
\usepackage{hyperref}
\usepackage[normalem]{ulem}
\usepackage{xcolor}

\usepackage[utf8]{inputenc}
\usepackage[acronym,style=super,nonumberlist,toc]{glossaries} 

\begin{document}
\title{Improved Limits on \texorpdfstring{\ensuremath{n \rightarrow n'}}{n n'} Transformation from the Spallation Neutron Source}
\author{Francisco M. Gonzalez}
\email{gonzalezfm@ornl.gov}
\affiliation{Oak Ridge National Laboratory, Oak Ridge, TN 37831, USA}
\author{Cary Rock}
\affiliation{University of Tennessee, Knoxville, TN 37996, USA}
\author{Leah J. Broussard}
\affiliation{Oak Ridge National Laboratory, Oak Ridge, TN 37831, USA}
\author{Lisa DeBeer-Schmitt}
\affiliation{Oak Ridge National Laboratory, Oak Ridge, TN 37831, USA}
\author{Matthew J. Frost}
\affiliation{Oak Ridge National Laboratory, Oak Ridge, TN 37831, USA}
\author{Lawrence Heilbronn}
\affiliation{University of Tennessee, Knoxville, TN 37996, USA}
\author{Erik B. Iverson}
\affiliation{Oak Ridge National Laboratory, Oak Ridge, TN 37831, USA}
\author{Yuri Kamyshkov}
\affiliation{University of Tennessee, Knoxville, TN 37996, USA}
\author{Michael Kline}
\affiliation{Department of Physics, The Ohio State University, Columbus, OH 43210, USA}
\author{David Milstead}
\affiliation{Department of Physics, Stockholm University, 106 91 Stockholm, Sweden}
\author{Devyn Powers}
\affiliation{University of Tennessee, Knoxville, TN 37996, USA}
\author{James Rogers}
\affiliation{University of Tennessee, Knoxville, TN 37996, USA}
\author{Valentina Santoro}
\affiliation{Department of Physics, Lund University, P.O. Box 118, SE-221 00 Lund, Sweden}
\affiliation{European Spallation Source ERIC, Partikelgatan 5, 22484 Lund, Sweden}
\author{Shaun Vavra}
\affiliation{University of Tennessee, Knoxville, TN 37996, USA}
\date{January 8, 2024}

\begin{abstract}

Conversions between neutrons $n$ and Dark Matter candidate sterile neutrons $n'$ have been proposed as a mechanism for Baryon Number $\mathcal{B}$ violation.
In the case that there is a small mass difference $\Delta{m}$ 
between the $n$ and the $n'$ states, oscillations can be induced by compensating for $\Delta{m}$ with a magnetic field.
A search for such neutron oscillations was performed at the Spallation Neutron Source by looking for anomalous neutron transmission through a strongly absorbing cadmium wafer inside of a $6.6$~T magnet. 
The approach described here saw no regenerated neutrons above background, which provides an improved limit for neutron - sterile neutrons transformations for a range of $\Delta{m}$ between $0.1$~neV and $1000$~neV. 

\end{abstract}
\maketitle

\section{Introduction}
\label{sec:intro}
\subsection{Motivation}
\label{ssec:motivation}

Much of the matter in the universe is invisible, not interacting via the electromagnetic force. 
This ``Dark Matter'' has thus far evaded direct detection, which has led to a robust suite of experimental and theoretical proposals looking for new exotic particles and interactions~\cite{Battaglieri:2017aum,Boddy:2022knd}. 
The traditional Cold Dark Matter paradigm, featuring non-relativistic matter interacting only by gravity, stands in tension with the observations of the density profiles of the dark matter halos of galaxies. 
One possible solution to this cusp/core problem introduces new forces acting between dark matter particles within a model of Self-Interacting Dark Matter~\cite{Tulin:2017ara,Adhikari:2022sbh}. 
Theories of Self-Interacting Dark Matter vary in complexity from introduction of exotic forces all the way up to atomic Dark Matter~\cite{Cyr-Racine:2012tfp}.

While the Dark Matter hypothesis can account for observations at the galactic scale, physical mysteries also appear at much smaller distances. The weak interaction in the Standard Model (SM) of particle physics is left-handed and maximally breaks parity. 
Early attempts to resolve this apparent asymmetry in chirality posited a right-handed equivalent of the Weak interaction and subsequently right-handed copies of the SM~\cite{Lee1957,Kobzarev:1966qya}. 
This right-handed ``Mirror Model'' (SM$'$) would completely duplicate the $SU(3) \times SU(2) \times U(1)$ form of the SM with a dark sector denoted as $SU'(3) \times SU'(2) \times U'(1)$~\cite{Foot:1991bp}. 

Mirror Matter, having the same interactions as the SM, could have similar complexity to observable matter on cosmological scales~\cite{Berezhiani:2003wj}. 
Any existing Mirror Matter would interact with normal matter predominately through gravity, making it a candidate for Self-Interacting Dark Matter. 
The added particles and forces would have implications for element formation in the early universe, and could even be a mechanism for a non-zero cosmological constant~\cite{Coc:2014gia,Cyr-Racine:2021oal}. 
This could potentially lead to the formation of Mirror Matter-based ``Dark Stars,'' or mixed stars with a dark matter subcomponent~\cite{Berezhiani:2005vv,Ciancarella:2020msu}. 
Such exotic objects could be sought with existing astrophysical dark matter searches~\cite{Howe:2021neq}. 

Laboratory searches for mirror dark matter probe the mixing mechanism of the neutrally charged components of the $SM$ and $SM'$ sectors, such as photons, neutrinos $\nu_{R}$, and particularly $n_{R}$ neutrons~\cite{Berezhiani:2005hv}. 
Such mixing leads to oscillations between the neutron $n$ and its sterile mirror partner $n'$, an effect that can be observed in laboratory experiments as a disappearance of a neutron with apparent violation of Baryon charge and unitarity. 
Outside of the Mirror Model paradigm, the search for $n \rightarrow n'$ transformation also has been proposed as a probe of multiple branes ~\cite{Sarrazin:2015sua}.

One of the major unsolved questions in physics is the universe's matter-antimatter asymmetry~\cite{Aghanim:2020}, which requires a violation of Baryon number $\mathcal{B}$~\cite{Sakharov:1967dj}. 
In a universe with Mirror Matter, SM neutral particles such as photons, the neutron $n$, or neutrinos $\nu$, could mix with their Mirror counterparts. 
Such mixing would cause $\mathcal{B}$ violation in the neutron sector or $\mathcal{L}$ violation in the neutrino sector. 
The neutron, due to its relatively long lifetime and relative ease of trapping, is a natural candidate for studying Mirror Matter in the laboratory.
Neutrons would oscillate into their Mirror counterpart $n'$, with some characteristic mixing mass denoted as $\epsilon_{nn'}$~\cite{Berezhiani:2005hv}. 
This mixing, as a $\Delta \mathcal{B} = 1$ process, can be contrasted with the well-studied neutron-antineutron $n \rightarrow \bar{n}$ oscillations, where $\Delta \mathcal{B} = 2$ for $n \rightarrow \bar{n}$~\cite{Proceedings:2020nzz, Addazi:2020nlz}. 
Mirror neutrons could also be an intermediary process to $n\rightarrow \bar{n}$ conversion, $n\rightarrow n'\rightarrow \bar{n}$, which remains a relatively unexplored parameter space~\cite{Berezhiani:2020vbe}. 

\subsection{Phenomenology of \texorpdfstring{$n$}{n} to \texorpdfstring{$n'$}{n'} Transformation}
\label{sec:phenomenology}
In the absence of mixing with mirror neutrons, the evolution of the non-relativistic neutron as it propagates in matter in the presence of an external magnetic field $\vec{B}$ and undergoes beta decay can be described by the non-Hermitian Hamiltonian (in natural units $c = \hbar = 1$), see e.g.~\cite{Kamyshkov:2021kzi}. 
\begin{equation}
    \label{eq:phen:hamiltonian}
    \mathcal{H}_{n} = m_n + \frac{p_n^2}{2 m_n} + \mu_n \vec{\sigma} \cdot \vec{B} + V - i\left(W + \frac{\Gamma}{2}\right)\,,
\end{equation}
where $m_n$ and $\mu_n$ are mass and magnetic moment of the neutron, $\vec{\sigma}$ is the vector of Pauli matrices, $V$ is the real and $W$ the imaginary absorptive parts of the optical potential, and  $\Gamma$ is the width of neutron decay. 
Equation (\ref{eq:phen:hamiltonian}) can be extended to the Mirror sector, where the mirror neutron mass $m_{n'}$, decay width $\Gamma'$, and magnetic dipole moment $\mu_{n'}$ are not necessarily the same as for the ordinary neutron. 
There could exist a mirror magnetic field $B'$ and a mirror matter induced spin-independent optical potential $V'-iW'$ \cite{Berezhiani:2009ldq}:
\begin{equation}
    \label{eq:phen:mirror_hamiltonian}
    \mathcal{H}_{n'} = m_{n'} + \frac{p_{n'}^2}{2 m_{n'}} + \mu_{n'} \vec{\sigma} \cdot \vec{B'} + V' - i\left(W' + \frac{\Gamma'}{2}\right)\,.
\end{equation}
The states of the neutron from Equation \ref{eq:phen:hamiltonian} and the mirror neutron from Equation \ref{eq:phen:mirror_hamiltonian} would exist independently if there is no mixing interaction leading to oscillation between these states. 

The $Z_2$ or $PZ_2$ (the latter with parity inversion) symmetry between ordinary and mirror sectors need not be exact. 
It can be broken spontaneously via a difference in the Higgs vacuum expectation values or due to the presence of new limited-range forces in the ordinary and mirror sectors. 
Such a difference would induce a small energy difference $\Delta{E}$ between the $n$ and the $n'$ states, or equivalently a mass difference $\Delta{m} = m_{n'} - m_{n}$. The value of $\Delta{m}$ can be either positive or negative, where by convention a positive sign for $\Delta{m}$ corresponds to $m_{n'} > m_{n}$. This paper considers both cases of $\Delta{m}$. 

Evolution of the two-state $(n,n')$ system follows the Schr\"{o}dinger equation $i \frac{d\Psi}{dt}=\mathcal{H}\Psi$, where $\Psi$ is the wavefunction of the $n$ and $n'$ components in two $(\pm)$ spin polarization states $\Psi=(\psi^{+}_n , \psi^{-}_n , \psi^{+}_{n'} , \psi^{-}_{n'})^T$. 
The transformational evolution of the $(n,n')$ system can then be described by introducing an unknown mixing parameter $\epsilon_{nn'}$ to the $4 \times 4$ Hamiltonian in the Schr\"{o}dinger equation: 
\begin{equation}
    \label{eq:phen:schrodinger}
    i \frac{d}{dt} \Psi = 
    \begin{pmatrix}
        \mathcal{H}_{n} && \epsilon_{nn'} \\
        \epsilon_{nn'} && \mathcal{H}_{n'}
    \end{pmatrix} \Psi\,.
\end{equation}
The mass mixing between the two states, $\epsilon_{nn'}$, can be alternately described using the characteristic oscillation time $\tau_{nn'} = \epsilon_{nn'}^{-1}$. 
In a more general case, the mixing element $\epsilon_{nn'}$ can contain a transition magnetic moment $\eta_{B} \vec{\sigma}$  or electric dipole moment $\eta_{E} \vec{\sigma}$, which could couple to the respective standard and mirror fields and modify the transition probability due to gradients in the magnetic or electric field~\cite{Berezhiani:2019qqw}. 
The search described in this paper assumes these $\eta_{B}$ and $\eta_{E}$ couplings are absent. 

Although Equation \ref{eq:phen:schrodinger} would generically describe the $(n,n')$ system, some simplifying assumptions can be made. When the momentum $\vec{p}_n$ is conserved, the difference in the kinetic energy terms in Equation \ref{eq:phen:schrodinger} due to the mass difference $\Delta{m}$ can be negligible. 
Symmetry breaking between ordinary and mirror sectors that would be responsible for a small $\Delta{m}$ between $n$ and $n'$ states might also induce $\mu_{n} \neq \mu_{n'}$.  
The magnetic field in the laboratory $B$ interacts only with the magnetic moment of the neutron $\mu_{n}$ and not with the magnetic moment of the mirror neutron $\mu_{n'}$, so a difference between the moments can be incorporated into the unobserved magnitude of $\vec{B'}$ or $\Delta{m}$. 
Although a mirror magnetic field $\vec{B'}$ as large as Earth's magnetic field cannot yet be ruled out in laboratory experiments~\cite{Berezhiani:2009ldq}, the dark matter density of the Earth suggests the mirror material potentials $V'$ and $W'$ to be vanishing~\cite{Ignatiev:2003js}. 
Since the time of flight of cold neutrons in the laboratory $\mathcal{O}(10^{-3}\,$s$)$ is significantly less than the neutron lifetime ($\tau_n\approx880$\,s), the decay rates $\Gamma$ and $\Gamma'$ are negligible. 
Thus, by assuming $V'=0$ and $W'=0$, and omitting the diagonal terms in the Hamiltonian that do not affect the oscillations, the $4 \times 4$ matrix Hamiltonian can be simplified to:
 \begin{equation}
\label{eqn:phen:simp_hamiltonian}
    \mathcal{H} = \left( \begin{array}{cc}
        U - i W& \epsilon_{nn'} \\
        \epsilon_{nn'} & 0
    \end{array} \right)\,,
\end{equation}
where $U$ includes only real numbers:
\begin{equation}
\label{eqn:phen:Udef}
U= V- \Delta{m} \pm \mu_{n} \vec{\sigma} \cdot \left(\vec{B} - \vec{B'} \right).
\end{equation}
This simplification is possible for either polarization of neutron in an unpolarized neutron beam if the transverse component of the magnetic field does not produce a spin rotation. 

The overall sign of $U$ depends on the value of $\Delta{m}$ as well as the surrounding materials. 
The sign of the magnetic field potential can be positive or negative depending on the polarization of the neutron and the direction of the magnetic fields $\vec{B}$ and $\vec{ B'}$. 
For most materials, the optical potential $V$ for a neutron interacting with matter is positive. Oscillations are driven by the off-diagonal mass mixing term $\epsilon_{nn'}$. The real diagonal elements of Equation~\ref{eqn:phen:simp_hamiltonian}, if larger than $\epsilon_{nn'}$, can suppress oscillations between $n$ and $n'$~\cite{Berezhiani:2009ldq}. 
Thus, an unknown $\Delta{m}$, $\vec{\mu}_{n'} \cdot \vec{B'}$ or any other environmental potential that is different for $n$ and $n'$ can reduce the oscillations induced by $\epsilon_{nn'}$. 
However, because these components could have different signs, it is possible to compensate for these competing potentials. 
The laboratory magnetic field $\vec{B}$ or the material optical potential $V$ can be tuned such that the sum of potentials vanishes, allowing oscillations to occur. 
Oscillations are maximal when both diagonal terms in Equation~\ref{eqn:phen:simp_hamiltonian} are zero. The evolution of the $(n,n')$ system depends upon the total difference in energy between the $n$ and $n'$,
\begin{equation}
    \label{eqn:phen:diff_energy}
    \Delta{E} = U - i W\,,
\end{equation}
where the imaginary part of the optical potential $W$ both causes attenuation of the $(n,n')$ system and affects the frequency of oscillations. 
The solution of the Schr\"{o}dinger equation describing oscillating ($n,n'$) in absorptive materials with a non-Hermitian Hamiltonian can be found in \cite{Kamyshkov:2021kzi}.

In vacuum with $V=0$ and $W=0$, and in the absence of fields which are different for $n$ and $n'$ such as the magnetic field $\vec{B}$, the time averaged probability of $n\rightarrow n'$ oscillation at small $\theta_{0}$ becomes~\cite{Berezhiani:2018eds}:
\begin{equation}
 \hspace*{-0.3cm}
    \label{eqn:phen:theta0}
P_{nn'}=\frac{2\epsilon_{nn'}^2}{\Delta{m}^2+4\epsilon_{nn'}^2}=\frac{1}{2}\sin^{2}{2 \theta_{0}} \simeq 2 \theta^2_{0} \simeq 2 \left( \frac{\epsilon_{nn'}}{\Delta{m}} \right)^2\,
\end{equation}
Thus, the probability $P_{nn'}$ in vacuum is determined by two independent parameters $\epsilon_{nn'}$ and $\Delta{m}$.  
The vacuum mixing angle $\theta_0$ defined by 
\begin{equation}
    \label{eqn:def:tan}
\tan{2 \theta_{0}} \equiv 2 \epsilon_{nn'} / \Delta{m}\,
\end{equation}
serves as a rotation angle for diagonalization of the Hamiltonian in Equation~\ref{eqn:phen:simp_hamiltonian}. When the angle $\theta_0$ is small, it becomes a convenient parameter for calculating the transformation probability $P_{nn'}$.  
$\Delta{m}$ and $\theta_0$ provide an alternative set of two independent parameters describing oscillation in a vacuum. 
Experimental limits below are presented in terms of the vacuum oscillation parameters $\Delta{m}$ and $\theta_0$.

From Equation~\ref{eqn:phen:theta0}, for free neutrons outside of any magnetic fields or material potentials, the oscillation probability is suppressed by $U=\pm \Delta{m}$.  
In the presence of a nonzero constant magnetic field $\vec{B}$, the angle required for diagonalization of the Hamiltonian, $\theta_{B}$, will be modified:
\begin{equation}
 \hspace*{-0.5cm}
    \label{eqn:phen:theta_B}
    \tan{2 \theta_{B}} =\frac {2\epsilon_{nn'}}{\Delta{m} \pm |\mu_{n} B|}=\frac{\tan{2 \theta_{0}}}{1 \pm \left|\mu_{n}B \right|/\Delta{m}}\,,
\end{equation}
where the $\pm$ indicates the possible presence of two polarizations of the neutron. 
At the magnetic field which compensates the mass splitting, $\Delta{m} - \mu_{n} B = 0$, there is a resonance where the oscillation is no longer suppressed. 
For an unpolarized beam of neutrons, half of the neutrons can reach this resonant condition. 

A $(n,n')$ system with a nonzero $\Delta{m}$ traveling through a non-uniform magnetic field passes through a dynamic resonance at $\Delta{m} - \mu_{n} B = 0$ when the mixing angle approaches its maximum value at $\theta_B = 45^{\circ}$. 
In this case, the transition probability will depend on $\epsilon_{nn'}$, on the neutron velocity $v$, and the gradient of magnetic field $\vec{B}$. 
This is a classical case of the Landau-Zener (LZ) non-adiabatic level-crossing transition~\cite{Landau:1932,Zener:1932ws}, which enhances the transformation $n \rightarrow n'$  in a similar manner to the solar flavor neutrino transitions~\cite{Kim:1987ss}. 
In the case of a beam of $n$ passing through a magnet with peak field $B_{max}$ and where $\Delta{m} < \left| \mu_{n} B_{max} \right|$, these LZ transitions occur at least twice; one ``level crossing'' occurs at the entrance of the magnet and one ``level crossing'' occurs at the exit of the magnet, when $\Delta{m}$ is compensated by $\mu_{n} B$. 
If the beam of neutrons is unpolarized, the polarity of the magnetic field does not matter as half of the neutrons will still undergo the LZ transition.
The amplitude of these transitions will depend on the initial phase of the oscillating $(n,n')$ system, the mixing $\epsilon_{nn'}$, the local gradient of magnetic field $\vec{B}$, and the velocity of the neutron $v$ as it passes through the ``level crossing.'' 
In the case where $\Delta{m} > \left| \mu_{n} B_{max} \right|$, the level-crossing does not take place and the transformation occurs adiabatically at $\theta_{0}<45^{\circ}$. 
In this case, the transformation probability is suppressed by $\Delta{E}=\Delta{m} - \left| \mu_{n} B_{max} \right|$ inside the magnetic field region and returns to the initial suppression defined by $\Delta{E}=\Delta{m}$ outside. 
An absorber inside the magnet will then ``measure" the system with the reduced suppression of probability.

Inside absorbing matter described by a complex optical potential $V - iW$, the oscillations between the neutron and mirror neutron states still continue with the Hamiltonian becoming non-Hermitian. 
The analytical solution of the Schr\"{o}dinger equation for the non-Hermitian Hamiltonian can be used to calculate the propagation of the oscillating $(n,n')$ system through a strong absorber where inside the absorber, $\theta_B$ becomes complex~\cite{Kamyshkov:2021kzi}: 
\begin{equation}
 \hspace*{-0.5cm}
    \label{eqn:phen:theta_B_gen}
     \tan{2 \theta_{B}} =\frac{\tan{2 \theta_{0}}}{\left( V-\Delta{m} \pm \left|\mu_{n}B \right| - iW\right)/\Delta{m}}.
\end{equation} 
The resonant condition in this case occurs when the real part $U = V \pm |\Delta{m}| \pm |\mu_{n}B|=0$. 

Analytic calculations of the evolution of the wave function using the Schr\"{o}dinger equation are difficult in the general case due to the requirement of matching the wave functions and their derivatives at both surfaces of the absorber as well as in the detector. 
Numeric calculations using density matrix evolution are free from these difficulties. 
The density matrix evolution equation or Liouville-von Neumann equation~\cite{isbn:978-0-19-852063-4} is a more general way of describing the interaction of a quantum system with its environment without using a wave function, and is equivalent to the Schr\"{o}dinger equation in case of a Hermitian Hamiltonian. 
Transmission probability calculations in this paper used the density matrix evolution equation with a Hamiltonian given by Equations \ref{eqn:phen:simp_hamiltonian} and \ref {eqn:phen:Udef} under the assumption of $B'=0$: 
\begin{equation}
\label{eq:phen:liouville_von_neumann}
    \frac{d \hat{\rho}}{d t} = -i \hat{\mathcal{H}}\hat{\rho} + i \hat{\rho} \hat{\mathcal{H}}^{\dagger} \,,
\end{equation}
where $\hat{\rho}(t) = \left|\Psi(t) \right> \left< \Psi(t)\right|$ is a $2 \times 2$ density matrix (assuming the $(n,n')$ state is not polarized). 
The diagonal elements of this matrix, $\rho_{11}(t)$ and $\rho_{22}(t)$, represent the probability of the oscillating system being found in either a pure $n$ or $n'$ state, respectively. 
The density matrix is Hermitian such that the off-diagonal elements satisfy $\rho_{12}^{\star} =\rho_{21}$. These latter elements contain information about the phases of oscillation. 

The Liouville-von Neumann equation (Equation~\ref{eq:phen:liouville_von_neumann}) does not treat decoherence effects, and thus, it does not account for the reset of oscillation phases when the $(n,n')$ system scatters on nuclei in materials. 
For experiments with a cold neutron beam, and for absorber layers which have a relatively small scattering probability, decoherence effects are non-essential since they involve only the neutron component of the mixed $(n,n')$ state and thus behave similarly to absorption by removing neutrons from the beam. 
More generally, however, instead of Equation~\ref{eq:phen:liouville_von_neumann} one can use the Lindblad equation for the density matrix evolution that includes the macroscopic effect of scattering-induced decoherence~\cite{isbn:978-0-19-852063-4}. 
This approach has been used e.g., by the STEREO Collaboration for the calculation of the probability of transformation of ``hidden" neutrons produced from neutrons in a research nuclear reactor through a thick layer of heavy water where the number of elastic collisions with the material is large~\cite{Almazan:2021fvo}.

\subsection{Current Limits on \texorpdfstring{$n$}{n} to Mirror \texorpdfstring{$n'$}{n'} Transformation} \label{ssec:other_experiments}

The existence of neutron - mirror neutron mixing would cause observable consequences which can be probed in the laboratory. 
Astrophysical constraints on $(n,n')$ mixing come from mass losses in pulsars and the temperatures of neutron stars~\cite{Goldman:2019dbq, McKeen:2021jbh, Goldman:2022rth}. 
New heavy particles that would contribute to the mechanism of $n \rightarrow n'$ transformation have not been observed at the Large Hadron Collider. 
The case of large $\Delta{m}$ or $\Delta{E}$ has also been considered~\cite{Fornal:2018eol, Berezhiani:2018udo, Sarrazin:2015sua}. Neutron - mirror neutron mixing would result in unexplained neutron disappearance in terrestrial cold and ultracold neutron measurements.  
Here, we discuss in detail specific limits from other measurements.

In ultracold neutron (UCN) material traps where the  wall optical potential is $V\sim100$\,neV, the lowest observed UCN loss factor per single wall collision was reported as $\sim 2 \cdot 10^{-6}$, higher than the expected values from calculation~\cite{Serebrov:2004zf, Arzumanov:2015tea, Serebrov:2017bzo}. 
Assuming that this loss can be ascribed to $n \rightarrow n'$ oscillations in the trap, this factor can be interpreted as a limit on the mixing angle $\theta_0 < 10^{-3}$~\cite{Berezhiani:2018eds}. 
However, this limit is valid only for $\Delta{m} < 2(V-\bar{T}_{kin})$, where $\bar{T}_{kin}$ is the average kinetic energy of UCNs in the trap and $V \sim 100$~neV. 
At $\Delta{m}$ above this limit, only one non-oscillating eigenstate can exist in the trap and thus the oscillation effect is absent.

For larger $\Delta{m}$, constraints can be obtained from observation of neutron propagation in weakly absorbing materials. 
The angle $\theta_0$ (Equation~\ref{eqn:def:tan}) has a maximum value of $\theta_{0}=45^{\circ}$, which can be reached for $\Delta{m}=0$ or for finite $\Delta{m}$ if $\epsilon_{nn'} \gg \Delta{m}$. 
The time averaged probability of $(n\rightarrow n')$ transformation in Equation~\ref{eqn:phen:theta0} is $\sin^{2}{2\theta_{0}}/2$, which changes from 0.02 to 0.5 for any angle $\theta_{0}>0.1$. 
Neutron disappearance at this level would be observable for example for neutrons propagating in the heavy water reflector surrounding the compact central core in research reactors \cite{ILL-Argonne}. 
Every elastic collision would lead to the collapse of the $(n,n')$ system, leading to the loss of neutrons with probability $>2 \%$. 
The observation \cite{ILL-Argonne} that the radial neutron flux distribution calculated with MCNP agrees with the measured flux excludes $\theta_{0}>0.1$ for all values of $\Delta{m}$. 

Experimental limits on the Mirror Matter Model rely upon specific assumptions about the nature of $\Delta{m}$ and $\vec{B'}$ ~\cite{Pokotilovski:2006gq} (and possibly on the existence of a neutron transition magnetic moment~\cite{Berezhiani:2019qqw}). 
The most stringent transition limits, $\tau_{nn'} > 448$\,s, ($90 \%$ C.L.) come from the assumption of the perfectly degenerate case, with $\Delta{m}=0$ and $\vec{B'}=0$ and for the laboratory magnetic field $\vec{B}$ close to zero ~\cite{Serebrov:2008her}. 
As the presence of a magnetic field $\vec{B} \neq 0$ would suppress the rate of oscillations, the laboratory field in this experiment was shielded below the level of the quasi-free condition $|\vec{B}| < 100$\,nT.

In other searches with ultra-cold neutrons (UCN), the presence of mirror magnetic field $\vec{B'}$ with unknown magnitude and direction was considered \cite{Berezhiani:2012rq,Berezhiani:2017jkn,nEDM:2020ekj}. 
The loss rate of neutrons inside the storage traps were measured as a function of a laboratory magnetic field which should compensate for the suppression that would be provided by a possible mirror magnetic field. 
The presence of a mirror magnetic field $\vec{B'}$ was probed by utilizing an equal strength magnetic field in two directions, $B_{\uparrow}$ and $B_{\downarrow}$ and measuring the asymmetry $A_{\uparrow \downarrow}$ in stored UCN~\cite{Berezhiani:2009ldq}, 
\begin{equation}
    \label{eq:exp:asymmetryB}
    A_{\uparrow \downarrow} = \frac{n(B_{\uparrow}) - n(B_{\downarrow})}{n(B_{\uparrow}) + n(B_{\downarrow})}.
\end{equation}
This asymmetry was measured for different field strengths and directions to search directly for the $\vec{B'}$ providing the $n \rightarrow n'$ transformation effect. 
An additional asymmetry~\cite{Berezhiani:2009ldq} $E_0$ can be produced in the zero field case:
\begin{equation}
    \label{eq:exp:asymmetry0}
    E_0 = \frac{2 n(B = 0)}{n(B_{\uparrow}) + n(B_{\downarrow})} - 1.
\end{equation}
A reanalysis of those previous experiments reported a result consistent with $n \rightarrow n'$ losses for nonzero $\vec{B'}$~\cite{Berezhiani:2012rq, Berezhiani:2017jkn}. 
Subsequent dedicated searches for $n \rightarrow n'$ utilized repurposed experiments to search for the neutron electric dipole moment, and have excluded most of these claimed signals~\cite{Altarev:2009tg,Berezhiani:2017jkn,nEDM:2020ekj}. 
More UCN storage search experiments are still in progress ~\cite{Ayres:2021zbh, Mohanmurthy:2022dbt}. 
In addition to using UCN storage, further $n \rightarrow n'$ disappearance searches have been performed using beams of UCN passing through a guide with a tunable magnetic field~\cite{SaenzArevalo:2022umy}, with results obtained using the GADGET UCN detector at the ILL~\cite{Ban:2023cja}.

Neutron regeneration experiments, which search for the double transition $n \rightarrow n' \rightarrow n $, have been discussed in~\cite{Berezhiani:2017azg, Schmidt2007}. 
Neutrons which disappear before passing through an absorber would be regenerated on the other side, analogous to similar experiments using photons through a wall to search for exotic electromagnetic couplings~\cite{Redondo:2010dp}. 

Searches for sterile ``hidden'' neutrons $n'$ performed at nuclear research reactors ~\cite{Stasser:2020jct, Almazan:2021fvo} have an advantage of using very large initial number of neutrons for conversion to $n'$. 
Thermal neutrons propagating in the material in a reactor, in particular in heavy water, experience a Fermi potential suppressing the transformation. 
Neutrons undergo a large number of collisions with the reactor material that ``measures" the $(n,n')$ quantum system, providing conversion to $n'$. 
For a non-degenerate $(n,n')$ system with $\Delta{m} \neq 0$, the additional suppression of the probability can be compensated by the Fermi potential of the material.  
Regeneration of $n'$ to observable $n$ occurs in the detector outside the reactor. 
Repurposed detectors from sterile neutrino experiments were used to place strong limits on $n$ conversion to sterile $n'$~\cite{Stasser:2020jct, Almazan:2021fvo}. 

Further searches for such processes have been proposed for a higher flux source with a lower background detector ~\cite{Hostert:2022ntu}. 
The effect of neutron regeneration would be seen in a detector as a $n$ counting rate signal above the background correlated with the operation of the source. 
With detailed accounting of the initial neutron flux of the source, as well as the materials or magnetic fields responsible for the $(n,n')$ conversion, the observed $n$ signal can be translated into a probability of $(n \rightarrow n')$ oscillation. 

The theoretical $n \rightarrow n'$ model with $\Delta{m} \neq 0$ probed in this paper was originally proposed in \citep{Berezhiani:2018eds} as a potential explanation for the ``neutron lifetime anomaly.'' 
The lifetime of UCN stored inside of a magnetic or material ``bottle'' has been measured as $\tau_n = 878.4 \pm 0.5$\,s~\cite{Serebrov:2004zf, Pichlmaier:2010zz, Steyerl:2012zz, Arzumanov:2015tea, Ezhov:2014tna, Serebrov:2017bzo, Pattie:2017vsj, UCNt:2021pcg}. 
This value disagrees with the lifetime of $\tau_n = 888.0 \pm 0.7$\,s determined by measuring neutron decay products in a cold neutron ``beam'' by $> 4 \sigma$~\cite{Nico:2004ie, Yue:2013qrc, Sumi:2021svn}\footnote{In addition to the ``neutron lifetime anomaly'' between ``beam'' and ``bottle'' experiments, a second $3 \sigma$ anomaly exists between experiments storing UCN in material bottles ($\tau_n = 880.0 \pm 0.7$\,s)~\cite{Arzumanov:2015tea,Serebrov:2017bzo}, and storing UCN in magnetic bottles ($\tau_n = 877.8 \pm 0.2$\,s)~\cite{Ezhov:2014tna,UCNt:2021pcg}. 
A $n \rightarrow n'$ transformation model with a transition magnetic moment $\eta_B$ could potentially resolve this second discrepancy.}. 
The paper \citep{Berezhiani:2018eds} hypothesized that the apparent increase of $\sim1$\% in the measured $\tau_n$ comes from an increased $n \rightarrow n'$ oscillation probability due to the magnetic fields $\vec{B}$ present in the most precise ``beam'' experiment~\cite{Nico:2004ie,Yue:2013qrc}. 
The size of the discrepancy would indicate a potentially higher mass splitting $\Delta{m}$ than probed in small $\vec{B'}$ search experiments.
An experimental probe of this effect was proposed at Oak Ridge National Laboratory (ORNL), utilizing the Spallation Neutron Source (SNS) and the High Flux Isotope Reactor (HFIR)~\cite{Broussard:2017yev, Broussard:2019tgw}. 
A first search performed at SNS ~\cite{Broussard:2021eyr} excluded $n \rightarrow n'$ transitions as an explanation for the neutron lifetime discrepancy. 
This paper improves upon the results of this experiment to further limit the observed amount of neutron-mirror neutron mixing. 

\section{Experimental Approach} \label{sec:experiment}
\begin{figure*}
    \centering
   \includegraphics[width=\textwidth]{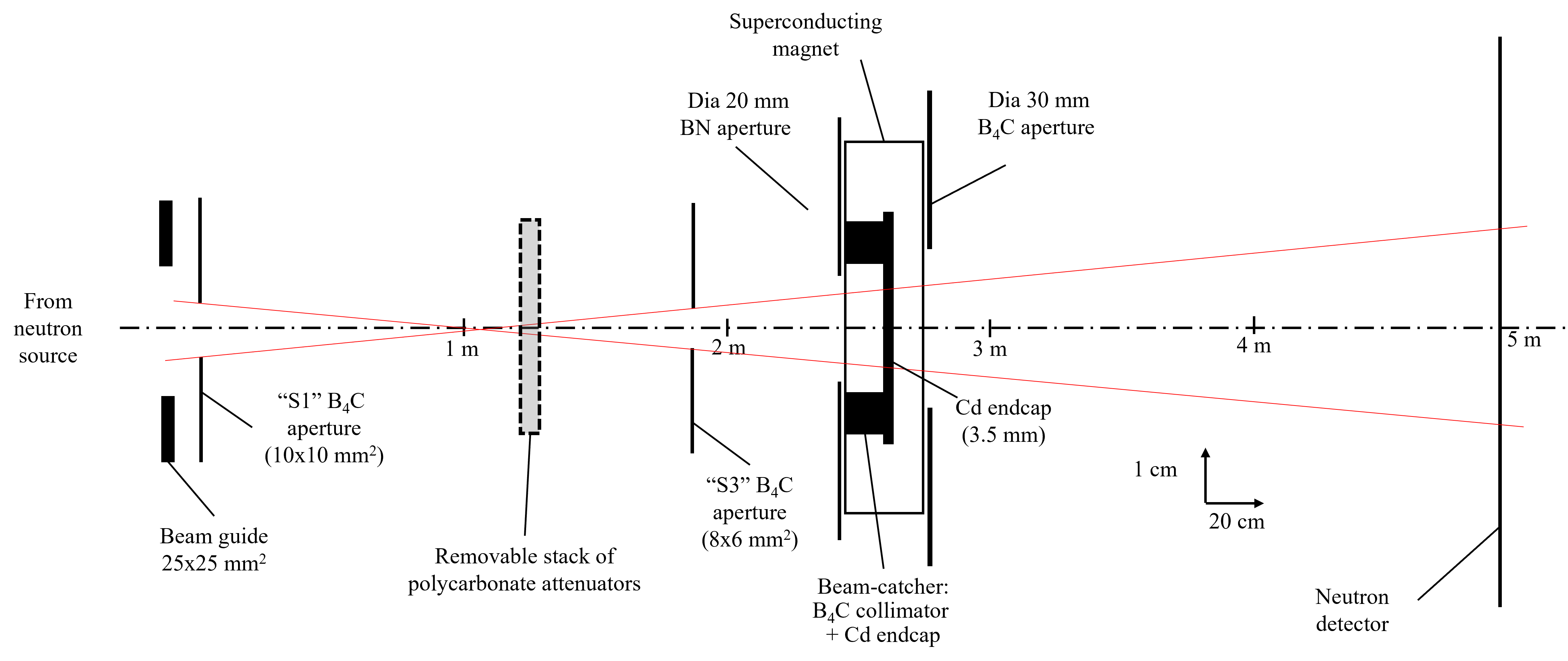}
    \caption{Diagram of the downstream portion of the MagRef beamline with the Cd absorber in place.}
    \label{fig:exp:magref_diagram}
\end{figure*}

The measurement reported here was performed using the cold neutron regeneration technique at the Magnetism Reflectometer (MagRef) instrument at the SNS at ORNL. 
This measurement implemented several improvements over the previous search reported in~\cite{Broussard:2021eyr}, including higher neutron intensity, improved intensity determination, longer experiment running time, more sensitive choice of neutron absorber material, and reduced backgrounds. 
A diagram of the MagRef beamline configuration in this experiment can be seen in Figure \ref{fig:exp:magref_diagram}. 
The major components of the experiment include the neutron source; beamline components including defining apertures, optional attenuators, and detector; and at the heart of the experiment, the magnet with neutron-absorbing beam-catcher. 

\subsection{Neutron Source}

The SNS produces H$^{-}$ ions which are accelerated by the linear accelerator to $1$\,GeV with a nominal power of $1.4$\,MW at the time of the measurement~\cite{Henderson:2014610}. 
The accelerated H$^{-}$ beam is stripped of electrons by a diamond foil, and then collected and intensified in the accumulator ring. 
The protons are then extracted in $700$\,ns long bunches with a repetition rate of $60$\,Hz which bombard a liquid mercury target to produce fast neutrons by spallation~\cite{Haines:201494}. 
The charge in each proton bunch is measured by an integrating current beam monitor with an accuracy of $3\%$~\citep{Blokland:2006dh}. 

\begin{figure}
    \centering
    \includegraphics[width=\columnwidth]{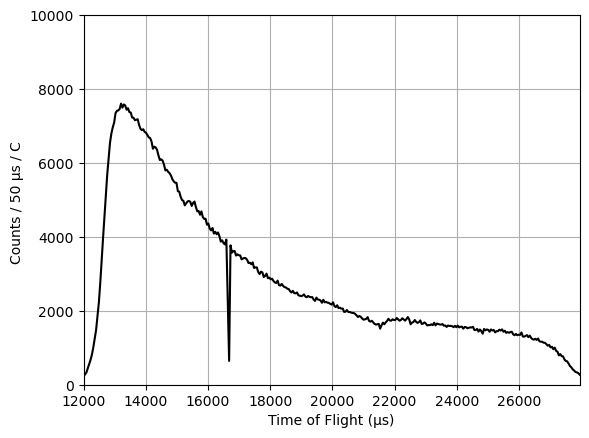}
    \caption{Time of flight spectrum for neutrons hitting the MagRef detector during a run taken with a small pinhole in the beam. This includes a $50$\,$\mu$s cut around $16666$\,$\mu$s to exclude the prompt flash.}
    \label{fig:results:time_of_flight}
\end{figure}

The neutrons viewed by the MagRef beamline~\citep{doi:10.1080/10448630802210537, LAUTER20092543} are moderated by a $20$\,K liquid hydrogen moderator~\cite{Lu:2008268}. 
Neutrons are emitted with a time scale of tens of $\mu$s and transported via a curved guide over a distance of approximately $16$\,m to the experimental area. 
A system of beam choppers allows the passage of neutrons only within a user-selected wavelength range, in this case from $2.2$ to $5.1$\,\AA. The neutrons then pass through the $5$\,m long experimental area before reaching the detector. 
The time at which they complete this flight, referenced to the clock signal corresponding to the arrival of protons on the spallation target, provides a Time of Flight (TOF) estimate for each neutron hitting the detector. 
A typical TOF spectrum of the neutron beam is shown in Figure \ref{fig:results:time_of_flight}. 
Determination of the absolute neutron intensity is discussed in the Results section. 
The neutrons passed by the chopper system arrive at the detector position at times ranging from $12$ to $27$\,ms following their emission from the moderator surface. 
The analysis discussed in this work removed neutrons near the edges of the chopper spectrum, only including neutrons in the interval of TOF between 13.0\,ms and 26.0\,ms. 
This TOF exceeds the time between accelerator frames at $60$~Hz, resulting in a ``prompt flash'' of background at $16666\,\mu$s. 
This background is excluded from further analysis by ignoring any events arriving in a $50\,\mu$s window around $16666\,\mu$s. 
At nominal SNS power of 1.4\,MW the proton beam charge was fairly stable with an average charge of $Q_1=23.34$\,$\mu$C per proton pulse. 
To avoid beam instabilities, which could affect the detector response, neutrons hitting the detector from frames with  $Q_1<23.0$\,$\mu$C were excluded from this analysis. 

The TOF can be related to the neutron wavelength by using the length of the beam path and calibrated using observed dips in the TOF spectrum caused by Bragg scattering of neutrons from aluminum at $2.864$, $4.050$ and $4.676$\,\AA. 
Since those dips occur at very specific and well defined wavelengths, they are suitable for converting time of flight in the detector to proper wavelength. 
This can be done by using the relationship:
\begin{equation}
    \label{eq:app:time_of_flight}
    \lambda(t)=3.956\frac{t}{L}
\end{equation}
Here, time $t$ is in $\mu$s and length $L$ is in mm. 
As seen in Figure \ref{fig:results:time_of_flight}, there are three distinct dips seen at $15230$, $21430$ and $24658$\,$\mu$s. 
Using these, one finds a distance $L$ to be $20955 \pm 11$\,millimeters, very close to the nominal quoted source-to-detector length of $21$\,m. 

\subsection{MagRef Beamline and Detector}

Upon exiting the neutron guide inside the biological shielding surrounding the SNS target monolith, the neutrons encounter a set of vertical and horizontal B$_4$C blades (``S1'' in Figure~\ref{fig:exp:magref_diagram}) serving as an aperture which allows a neutron beam with a maximum permitted extent of $\sim30$\,mm$\times 30$\,mm to enter the MagRef experimental area. 
The neutrons travel unimpeded through air nearly $2$\,m before encountering a second aperture which defines the divergence of the beam. 
For this measurement, the upstream aperture (S1) permitted a beam extent of 10$\times$10\,mm$^2$ and the downstream aperture (S3) permitted a beam extent of 8$\times$6\,mm$^2$ (horizontal $\times$ vertical). 
The neutron beam divergence was checked by comparing the  neutron beam extent measured by the main MagRef detector to the extent measured using a neutron camera placed about 2\,m upstream. 
Simulation studies confirmed that the neutron beam divergence did not significantly impact the sensitivity of this measurement. 

A set of commercial polycarbonate (PC)  $(C_{16}O_{3}H_{14})_{n}$ plates stacked together was installed 0.7\,m upstream of the downstream aperture (S3) and used to attenuate the beam for an indirect determination of the total neutron intensity, as described in the Results section. 
It was removed during all other measurements. 
The stack of plates was aligned against a square edge to ensure that the faces of the plates were perpendicular to the beam. 
The nominal ``sample position'' of the beamline, where in typical material science measurements neutrons are scattered from material samples inside of a magnet, is 0.7\,m past the downstream aperture (S3). 
In the measurements reported here, a 3.5\,mm thick absorbing Cd plate was installed at the sample position. 
The magnet configuration used for this experiment is described in the next subsection. 
Downstream, the neutron beam passes through a shielded beam tube before encountering the detector 2.3\,m downstream of the sample position. 
A final set of vertical and horizontal B$_4$C blades in front of the detector were positioned to fully enclose the shielded beam tube to reduce scattered neutron backgrounds from the room. 
Although MagRef has the capability to polarize the neutron beam, these measurements did not use a polarized beam. 

The MagRef detector was developed by Brookhaven National Laboratory and uses a $^3$He converter to detect neutrons through the reaction $n + ^3He \rightarrow p + t$ with $>90\%$ efficiency. 
The $p$ and $t$ have a short range $O$(mm) in the nominal 6.8\,atm of $^3$He plus 2.7\,atm propane (buffer gas) in which they produce ionization. 
The multi-wire proportional counter detects the ions, and provides position sensitivity through 
cathode readout with multi-node charge division with spatial resolution of about 1.4\,mm. 
The detection efficiency is not perfectly uniform with position, but the technique reported here is not sensitive to nonuniformities in efficiency. 

An advantage of this detector technology is the ability to discriminate against gamma-rays, due to the much lower ionization density of gamma induced electrons. 
This is critical given the use of a fully-absorbing Cd sheet required for this experiment which fully converts the neutron flux at the sample position to an isotropic gamma field. 
The detector is installed on a rotating arm, used to view large scattering angles in other experiments, but was oriented perpendicular to the beam axis for these measurements. 

Data taken using the detector is represented as an array of $304 \times 256$ ($X \times Y)$ pixels, each with area $0.7 \times 0.7$\,mm$^{2}$~\cite{RADEKA1998642}. 
The maximum total counting rate of the detector is limited by the readout dead time of $4$\,$\mu$s. 
Each single detected neutron is subsequently recorded into a DAQ system which includes a GPS-based time tag, the $(X,Y)$ position of the pixel in the detector, the time-of-flight, and the proton charge for the SNS pulse that produced the neutron. 

\begin{figure}
    \centering
    \includegraphics[width=\columnwidth]{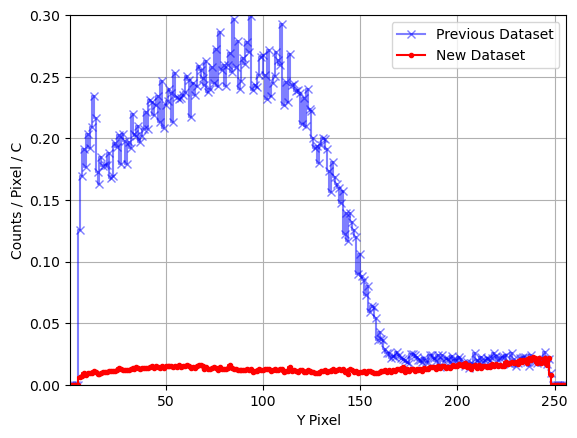}
    \caption{Background counting rate normalized to proton charge versus vertical position on the detector during the measurement reported in \citep{Broussard:2021eyr} (blue) and after improving shielding against backgrounds for the measurement reported here (red). (Color online)}
    \label{fig:exp:bkgdreduct}
\end{figure}

The ambient background counting rate was a significant limiting factor in the previous measurement reported in~\cite{Broussard:2021eyr}. 
For this measurement, the detector was surrounded by additional shielding made of boron carbide and $\sim$5\% (by mass) borated polyethylene. 
The downstream neutron beam guide was brought closer to the detector apertures. This removed a line of sight to neutrons scattering from the granite and concrete floor. 
The background counting rate was reduced by a factor of ten and a significant asymmetry in the vertical direction was eliminated (Figure~\ref{fig:exp:bkgdreduct}). 
Measurements performed with the neighboring beamline shutter opened and closed confirmed that no additional background was observed due to external sources. 

\begin{figure}
    \centering
    \includegraphics[width=\columnwidth]{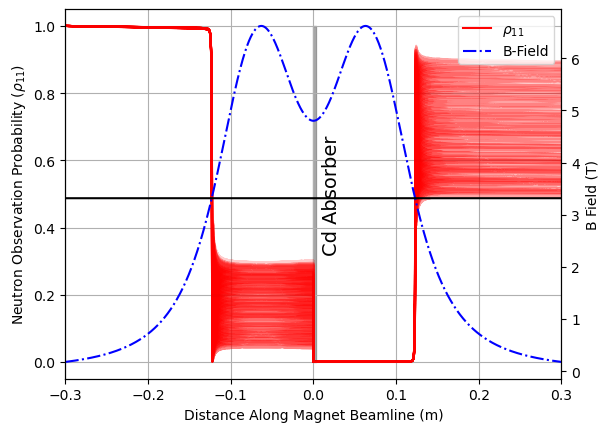}
    \caption{Example of a calculated evolution of the $(n,n')$ system through the MagRef field and absorber for $\Delta{m} = 200$\,neV, $\theta_0 = 5\times10^{-3}$ and using 100 wavelengths sampled from the distribution in Figure \ref{fig:results:time_of_flight}. The black horizontal line indicates the magnetic field that would correspond to $\Delta{m}$. The square of the neutron wave function as a probability of neutron observation is shown for each velocity sampled from the distribution. The effect is calculated for an absorber of $3.5$\,mm Cd (grey vertical line) as used in this experiment (color online).}
    \label{fig:exp:magneticField}
\end{figure}

\subsection{Magnet and Beam-Catcher}

The central feature of the experimental approach is the neutron evolution through the magnetic and material optical potentials which create the disappearance and regeneration effect, depicted in Figure~\ref{fig:exp:magneticField}. 
Example neutron probabilities through the magnetic field are shown. 
With no magnetic field, neutrons are fully absorbed in the material, and none are detected by the MagRef detector. 
In the presence of the magnetic field, the LZ transition enhances the probability that the neutron is ``observed'' in the mirror neutron state at the absorber, and another LZ transition results in a regeneration of neutrons upon exiting the magnet.

For this experiment, a Cryomagnetics, Inc.\ superconducting split pair magnet was installed in the MagRef beamline. 
The magnet is 30\,cm long and composed of two solenoids, such that the magnetic field has a double-peak~\cite{Magnet}. 
The dip in the center of the magnetic field is caused by a gap between the solenoidal coils intended for samples to be inserted for exposure to the neutron beam. 
When fully ramped, the magnetic field in the center of either of these two solenoids (at the local maxima at $6.3\,$cm upstream and downstream of magnet's center) is $6.6$\,T, while the field in the dip at the center of the magnet is $5$\,T. 
The magnet vendor provided the calculated field profile for radii up to 5\,cm and for $\pm$30\,cm along the magnet axis.  
The measured stray fields outside the magnet were also provided. Some components of the MagRef beamline increased the magnetic field locally. 
The ambient magnetic field at the beamline entrance at slit S1 was measured to be 10\,G and the field at the detector was less than 1\,G, which sets the limits of our sensitivity. 
From simulation, the experiment sensitivity was not significantly impacted by the shape of the magnetic field in the beamline far from the magnet, as long as the neutron originated in a lower magnetic field. 

The magnetic field profile for the major component of magnetic field can be seen in Figure~\ref{fig:exp:magneticField}. 
The oscillation probability of the $(n, n')$ state depends on the magnetic field, the orientation of the neutron spin, the neutron velocity, and the parameters of the mirror matter model. 
The search was conducted with the center of the magnetic field set to $\pm 4.7$\,T and $\pm 2.35$\,T, as well as a background measurement with no magnetic field. 
The beam is unpolarized, and so the oscillation probability is independent of the polarity of the magnetic field. 
During steady state operation, the current supplied to the magnet was monitored and the fluctuations in the magnetic field ($<1\%$) were negligible for this measurement. 

Inside the magnet, a 15\,cm long and 25.4\,mm inner diameter neutron-absorbing cylinder (B$_4$C) with an absorbing end-cap (cadmium) was installed in the beam path to remove neutrons while allowing $n'$ transmission, serving as a beam-catcher. 
A tightly-fitting 3D-printed Tough PLA (polylactic acid) support was used to axially center the beam-catcher in the magnet, with a slight ($<$1~mm) vertical displacement due to gravity.  
The end-cap was installed centered inside the magnet with a positioning precision of about 1~mm. 
The neutron absorber end-cap was changed from $32$~mm of B$_4$C as used in the previous setup \citep{Broussard:2021eyr} to $3.5$~mm of cadmium (Cd) for better sensitivity as described later. 

The intent of the beam-catcher is that the neutrons enter the tube unimpeded and are absorbed by the end-cap, but if neutrons backscatter they can be absorbed by the walls instead of potentially increasing backgrounds to the detector. 
It was important to ensure that neutrons are not absorbed by the beam-catcher walls before reaching the end-cap, which would result in an overestimate of the neutron flux (and thus sensitivity) during the experiment. 
A neutron camera placed in the magnet position was used to estimate the optimal upstream aperture settings in order to maximize the intensity through the beam-catcher without any loss or scattering of neutrons in the beam-catcher walls. 
The magnet was installed on a 2-dimensional linear and rotational stage, and the position and vertical axis angle were scanned to find the peak count rate on the detector, such that the neutron beam was roughly centered. 
To ensure the neutron beam did not encounter the beam-catcher walls, a 20\,mm boron nitride aperture was installed at the front face of the magnet and a 30\,mm B$_4$C aperture was installed on the downstream face of the magnet to prevent scattering outside of the beam-catcher. 
The beam-catcher was centered in the magnet axis and the aperture was centered on the magnet with an accuracy of $<1$~mm. 
A comparison of the neutron beam position distribution as measured by the MagRef detector using different apertures demonstrated that the extent of the neutron beam was centered inside the beam-catcher sufficiently far from the beam-catcher walls. 
\section{Results} \label{sec:results}
\subsection{Intensity Measurement Analysis} \label{ssec:intensity}
The probability of the $n \rightarrow n'$ effect probed here must be determined for each initial wavelength present in the beam. 
Thus, the absolute spectral intensity of the cold neutron beam must be determined. 
The total neutron intensity reaching the MagRef detector varies with SNS proton beam charge, the conditions of the moderator, and the neutron beamline settings. 
While the proton charge is monitored and recorded for each SNS 60\,Hz pulse, the moderator condition can vary slowly with an impact on intensity that is not tracked. 
Further, beamline settings such as aperture positions and sample locations are nominal only. 
Therefore, the spectral intensity of the neutron beam is not necessarily constant and must be determined for each experiment configuration separately, with a correction for the proton beam charge. 
As the MagRef beamline was not equipped with any additional device for intensity measurement, the spectral intensity was determined indirectly using the MagRef detector. 

\begin{figure}
    \centering
    \includegraphics[width=\columnwidth]{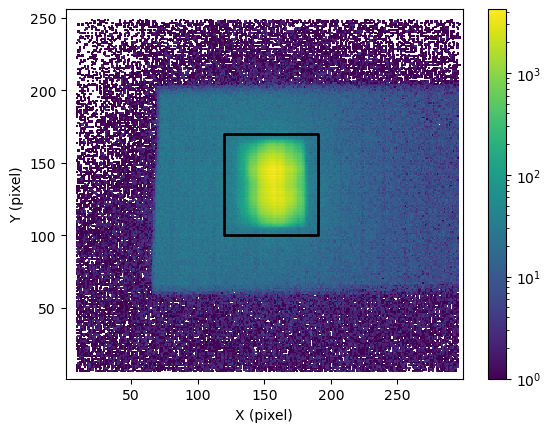}
    \caption{2D plot illustrating intensity calibration run taken with a stack of 18 PC plates     upstream of the MagRef detector. The attenuated primary neutron beam can be clearly seen as the peak inside the black square indicating the region of interest (ROI).  The shielding around the detector causes a sharp falloff in the background counting rate.
(color online).}
    \label{fig:results:beam_atten_detect}
\end{figure}

The full beam intensity needed for the experiment was too high for the MagRef detector to be measured without saturation, excess noise, or damage to the detector. 
The determination of the intensity of the neutron beam was therefore performed using an indirect approach, by measuring the intensity of the beam after passing through scattering attenuators with different thicknesses. 
The attenuator was composed of a variable number of PC plates with equal thickness. 
The thickness of each plate was measured to be $0.127$\,cm and the density was $1.195$\,g/cm$^3$. 
Six configurations using stacks of 18, 19, 20, 21, 22, and 24 PC plates were positioned in the beam $3$~m upstream from the detector for 20 minutes of data collection in each configuration. 
The maximum detector counting rate during the intensity calibration was $\sim2500$\,cps with 18 PC plates stacked together.  
A typical beam profile with 18 PC plates can be seen as a 2D plot in Figure~\ref{fig:results:beam_atten_detect}. 

The MagRef detector counting rates (per Coulomb of the proton beam charge) were measured for different configurations with different total PC thickness, then extrapolated to the counting rate of zero PC thickness to determine the original beam intensity. 
The neutron beam attenuation in the PC material is mostly due to elastic scattering on hydrogen. When combined with attenuation of the non-scattered beam and some small absorption, this scattering generated background around the primary beam peak in the detector.

The primary beam was centered at pixels (X,Y) = (156, 136) of the MagRef detector with full width half maximum (FWHM) in X and Y of 24 pixels and 41 pixels, respectively, as shown in Figure \ref{fig:results:beam_atten_detect}. 
The beam spot forms a peak which corresponds to neutrons passing through the PC attenuator stack without interaction. 
The direct beam-on area in the figure was used to determine a Region of Interest (ROI) of $70 \times 70$ pixels, which was used in searching for the effect of $n\rightarrow n'$. 
Outside the ROI in the detector, there is a background generated mostly by the neutrons scattered from the PC attenuator plates. 
The background falloff above pixel $Y=200$, below pixel $Y=70$, and on the left side of the detector below pixel $X\sim75$, is due to additional absorption provided by a shielding box around the beamline (See Figure \ref{fig:results:beam_atten_detect}). 
The region within these cut-offs is defined as the region of background (ROB). 
The background present in the ROI needs to be subtracted to determine the attenuated fraction of the beam that does not interact with the PC material. 

Two components in the measured background were identified. One component was almost uniform in the ROB of the detector as can be seen in Figure \ref{fig:results:beam_atten_detect}. 
This component can be described as a constant at some initial point with two slopes defining the background plane. The second component forms a scattering bump around the beam peak center. 
The shape of this bump is determined by the acceptance of the rectangular aperture S3 and the round 20-mm diameter BN apertures shown in Figure \ref{fig:exp:magref_diagram}. 
The bump is described by a 2D Gaussian around the center of the beam peak with three parameters: the height at the peak center and the two Gaussian widths $\sigma_{x}$ and $\sigma_{y}$. 
The description of these two components of background was confirmed by a PHITS \cite{Sato:2018imy} simulation using ENDF/B-VIII.0 data \cite{Brown:2018jhj} for a cold neutron beam scattering from PC material \footnote{The scattering kernel ``h-luci.40t'' for hydrogen bonded in PMMA that was used in PHITS came from the JAEA's MCNP-formatted JENDL-5 ACE library \cite{JENDL-5-ACE-library}, which is derived from the ENDF/B-VIII.0 data.} in the detection geometry shown in Figure \ref{fig:exp:magref_diagram}.

\begin{figure}
    \centering   
    \includegraphics[width=\columnwidth,trim=0.1cm 0cm 0cm 0cm,clip]{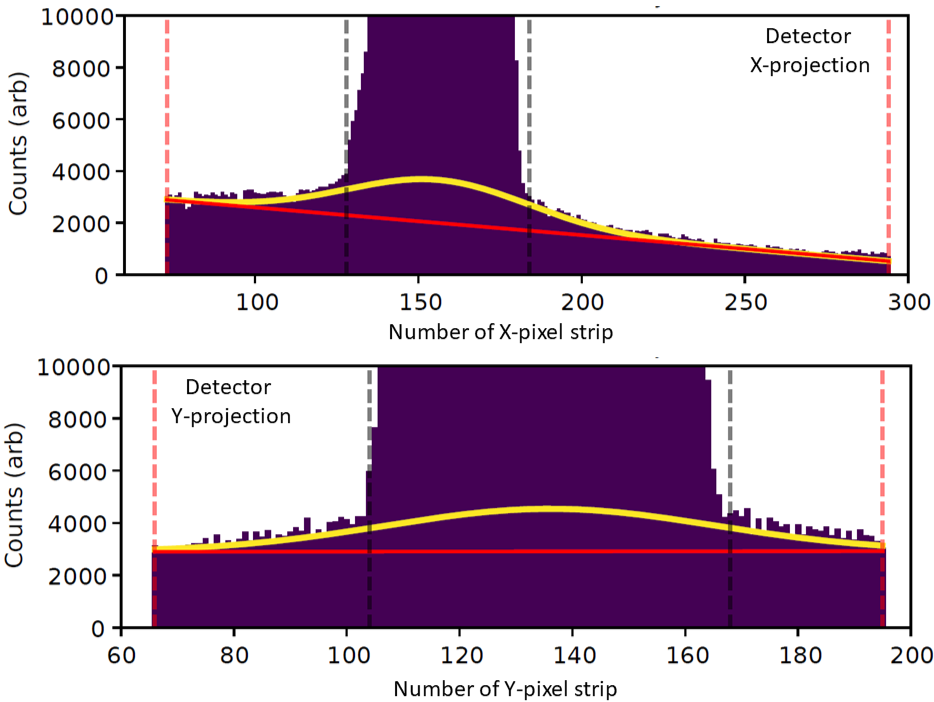}
    \caption{Example of a 2D background fit with 4 free parameters projected in X and Y for 18 PC attenuators (see text). Red vertical dashed lines correspond to the region of background (ROB) and black vertical dashed lines to the region of peak (ROP) which is excluded from the background fit. The solid lines show the contribution of the ``plane" background (red) and the ``Gaussian bump" background (yellow). Combined statistical and systematic uncertainties of the subtractable background determination are within the width of the Gaussian (yellow) line. In this example $\chi^2$/DOF=1.23.~(Color online)}
    \label{fig:results:bckgr_fit}
\end{figure}

The background observed in each PC measurement in the 2D matrix of pixels was fit to a 2D function combining the plane background with three parameters and the Gaussian bump background described with another three parameters. 
This functional form describes the general trends of the background rather than the detailed features in its structure. 
The region of the fit for minimization of $\chi^2$ included the ROB with the exclusion of a region containing the beam peak (ROP). 
The ROP used for the background fit was smaller than the ROI described above. 
The background fit regions and the number of background fit parameters were varied, leading to six different unique schemes for determining and subtracting the background in the region of the fit. These produced consistent results. 
The average typical $\chi^2$/DOF for all measured PC sets for the preferred subtraction scheme (to be discussed below) was $\sim$1.23 for 20,247 pixels. 
A typical example of a background fit in X- and Y-projections for 18-PC is show in Figure \ref{fig:results:bckgr_fit}. 

The values of the parameters obtained in the minimization of $\chi^2$ were used for the calculation of background by integrating the fitted function over the area of all pixels in the ROI. 
The ROI subtractable background is smoothly dependent on the number of PC plates and varies from $\sim 5\%$ for 18 PC plates to $\sim 18\%$ for 24 PC plates. 
All parameters are smoothly dependent on the number of PC attenuating plates. 
The Gaussian widths, $\sigma_{x}$ and $\sigma_{y}$, were consistent across all numbers of PC plates due to being determined solely by the acceptance geometry. 
For the final background subtraction scheme, the parameters $\sigma_{x}$ and $\sigma_{y}$ were thus fixed to their best fitted values to be the same for all PC runs. 
Thus, the number of free parameters for the background description is effectively reduced to four. 

The set of six measured attenuated beam intensities for 18, 19, 20, 21, 22, and 24 PC plates was obtained by subtracting the fitted background integrated over the ROI area from the total measured counts in the ROI for each of the six PC measurements. 
A detector dead time correction was applied to the measured counts.
A normalization for the variation of the SNS proton charge per run was also implemented to express measured intensities in counts per Coulomb. 
The uncertainties of the measured intensities included statistical error from the total counts in the ROI and the systematic error of the fitted background subtraction. 
The latter was estimated from each of the six fit covariance matrices by generating $10^{6}$  random samples of the background in the ROI with the background parameters (6 or 4) randomly taken from the multinormal distribution~\cite{multinormal} with the given fit covariance matrix. 
The standard deviation of the $10^{6}$ generated background values for every PC point was taken as a systematic error of the subtractable background in ROI. 
The same procedure was used for each of the different background subtraction schemes. 
The relative statistical error for the six measured points increases monotonically with the number of PC plates (18 to 24) from 0.1\% to 0.37\% and the relative systematic errors increases monotonically from 0.1\% to 0.86\%. 

For all subtraction schemes the six intensities measured with different number of PC plates were fitted to zero PC thickness with the function $I(k)=I(0)\cdot\eta^k$ with two fit parameters, where $I(0)$ is beam intensity for $k=0$ PC plates, and $\eta$ is the average beam attenuation factor per single plate. 
Statistical and systematic errors were added in quadrature as the errors used for the intensities in the fit procedure. 
The values of parameters $I(0)$ and $\eta$ for all background subtraction schemes agreed within uncertainties. 
This work uses the subtraction scheme with 4 background fit parameters, which provides slightly lower uncertainty for $I(0)$ and $\eta$ in the intensity fit. 
The results of the fit of six points extrapolated to the intensity with no attenuation for this background scheme resulted in the following values of parameters: $I(0)=(2.536 \pm 0.034)\times 10^{9} n/\text{C}$ and $\eta=0.6663 \pm 0.0006$. 
The $\chi^2$/DOF for these fits was relatively large at 6.46. 
However, the intensity found with 21 PC plates was contributing the most to the $\chi^2$/DOF. 
The 5-point fit without the intensity from the 21 PC plates measurement resulted in a reduced $\chi^2$/DOF to 2.14 with minimal change to I(0) and $\eta$: $I(0) = (2.466 \pm 0.034)\times 10^9$~counts$/$C and $\eta = (0.6673 \pm 0.0006)$. 
This could indicate an unaccounted systematic in the measurement using the 21 PC plates, such as an improper installation of the stack. 
Since the two intensities are consistent, we choose to use the intensity excluding the 21 PC plates, which corresponds to a lower overall sensitivity to the $n\rightarrow n'$ effect: 
\begin{equation}
    \label{eq:res:intensityfit}
    \begin{split}
    I(0)&=(2.466 \pm 0.034)\times 10^{9} \text{counts}/\text{C}\\
    \eta&=0.6673 \pm 0.0006
    \end{split}
\end{equation}

\begin{figure}
    \centering   
    \includegraphics[width=\columnwidth,trim=0.4cm 0.4cm 0cm 0.8cm,clip]{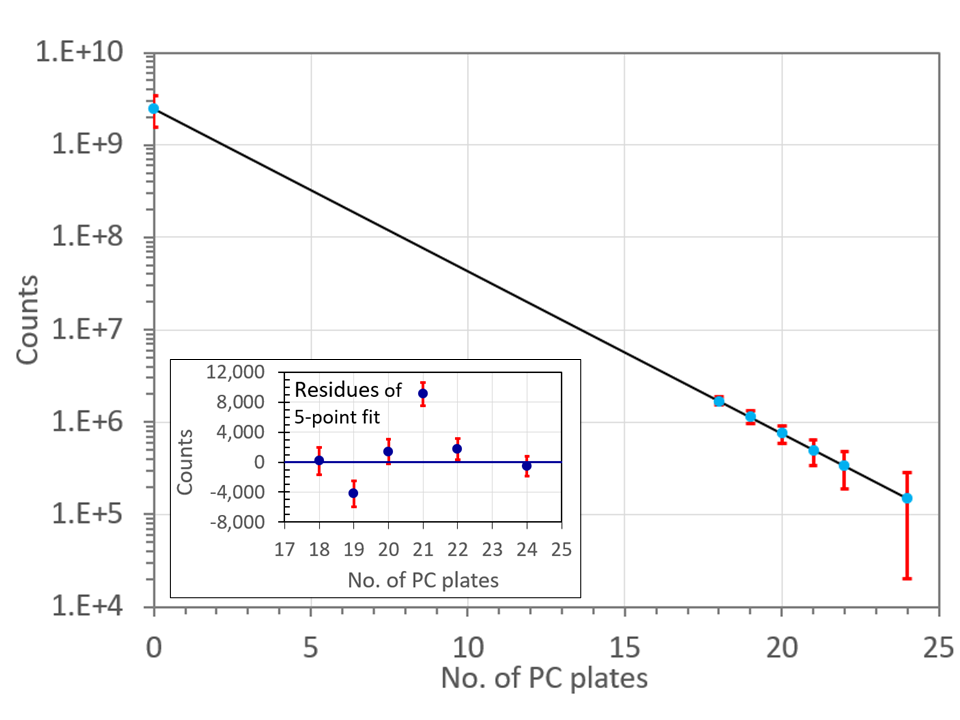}
    \caption{The extrapolation of the 5-point fit of measured intensities to zero PC plate thickness. The point for the 21 PC plates measurement is also shown in the plot for comparison. The six measured point errors are multiplied by factor 100, and the error $I(0)$ by factor 20. In insert: residues for the 5-point fit with actual errors. The sixth point for 21 plates is shown for illustration.(Color online)}
    \label{fig:5-point-fit}
\end{figure}

Figure \ref{fig:5-point-fit} shows the extrapolation of the 5-point fit to zero PC plate thickness. 
The point corresponding to the 21 PC measurement is also shown in the plot for comparison but was not used in the fit. 
The total errors shown in the plot for the six measured points are multiplied by a factor 100, and the error of the parameter $I(0)$ at zero thickness is multiplied by a factor 20 for visibility. 
The insert in Figure \ref{fig:5-point-fit} shows the residues for the 5-point fit using actual errors. 
The sixth point for 21 PC plates is shown for illustration. 
The total neutron beam intensity for the measurements reported here were somewhat higher than in the previous search~\cite{Broussard:2021eyr} due to the difference in slit configurations. 

\begin{figure}
    \centering   
    \includegraphics[width=\columnwidth,trim=0.1cm 0.8cm 0cm 1.3cm,clip]{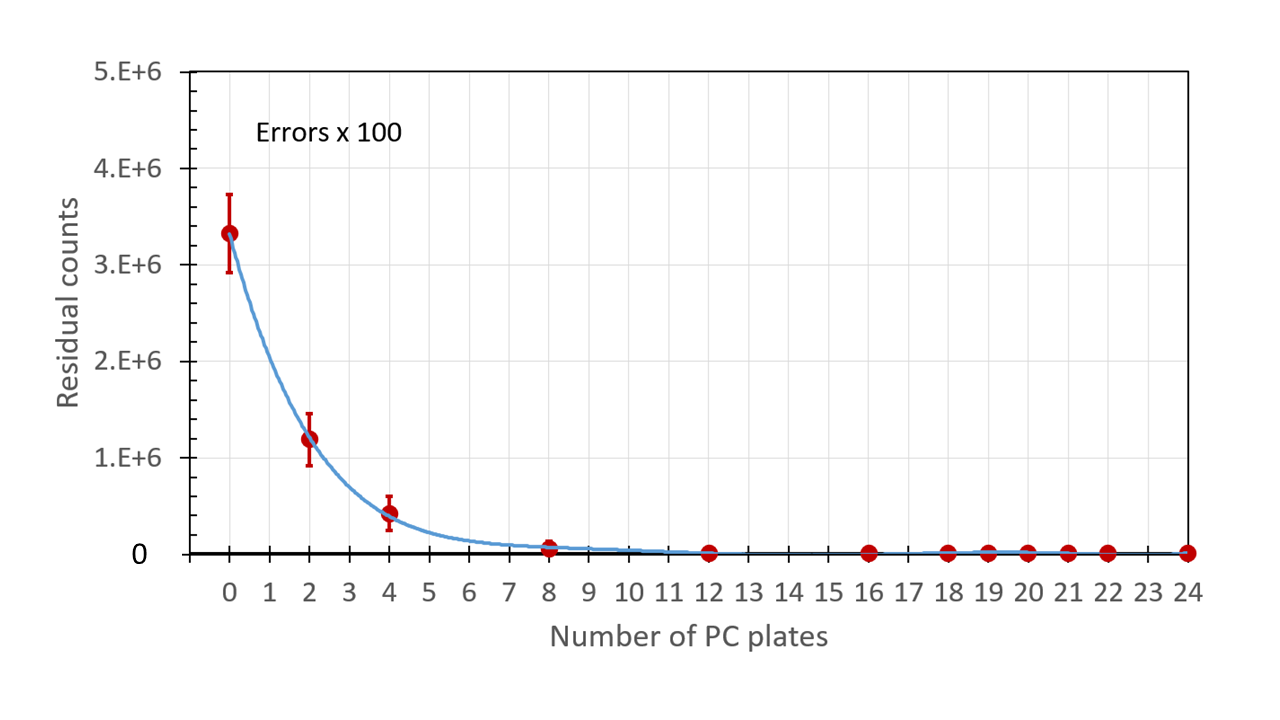}
    \caption{Residuals of a 2-parameter exponential fit of the last six intensity calibrating points corresponding to 18-24 PC attenuating plates, simulated with PHITS using room temperature PC material. The increase of the slope towards zero PC thickness demonstrates thermalization of the cold neutron spectrum, resulting in a required correction to the intensity measurements.(Color online)}
    \label{fig:Residues-of-fit}
\end{figure}

A PHITS simulation was used to estimate $\eta$ by using the scattering kernel for the bound molecular state of the scattering PC material. 
This simulation used the same attenuation thicknesses for each PC plate. The simulated number of neutrons transmitted to the detector without interaction for different PC thicknesses was fitted to extrapolate to the intensity of zero PC thickness, obtaining $\eta=0.6746 \pm 0.0009$, which is much closer to the experimental value of $\eta$ in (\ref{eq:res:intensityfit}). 
The disagreement with the measured $\eta$ is reasonable considering the uncertainty of the exact chemical structure of the PC material used, and the absence in the ENDF/B database of the scattering kernel for polycarbonate proper, where only the kernel for hydrogen in PMMA was available and used instead. 

Due to the very long extrapolation distance from 18 PC to 0 PC, the accuracy of the simple exponential behaviour was checked in simulation. 
An effect was observed due to up-scattering of the cold beam on the room-temperature material of PC plates in the PHITS simulation, by simulating smaller thicknesses of PC plates in the range from 0 to 18 plates. The effect of the increase of the local exponential slope (smaller $k$) is shown in Figure \ref{fig:Residues-of-fit}. 
In the simulation, the cold spectrum of neutrons scattered from the PC plates was transformed by scattering to a thermal neutron spectrum, which has been previously observed~\cite{Do:201442}. To account for this effect, a correction was applied to the experimentally determined zero PC intensity. 
The correction was taken from the PHITS simulation as a ratio of the intensity hitting the detector with zero PC attenuation, to the intensity determined by extrapolating to zero thickness from the set of simulated intensities attenuated by 18-24 PC plates. 
The correction factor was found as 1.254 $\pm$ 0.029. 
The same correction factor was applied to the experimentally measured extrapolated to zero PC intensity in Equation \ref{eq:res:intensityfit}, thus arriving at the final value of the intensity estimate in the experiment with relative error $2.7\%$:

\begin{equation}
    \label{eq:res:finalinten}
    I_{final}(0)=(3.092 \pm 0.083)\times 10^{9} \text{counts}/\text{C}  
\end{equation}
The uncertainty in the rethermalization correction factor dominates the determination of the beam intensity.

\subsection{Search for the Effect} \label{ssec:effect}

\begin{figure}
    \centering
    \includegraphics[width=\columnwidth]{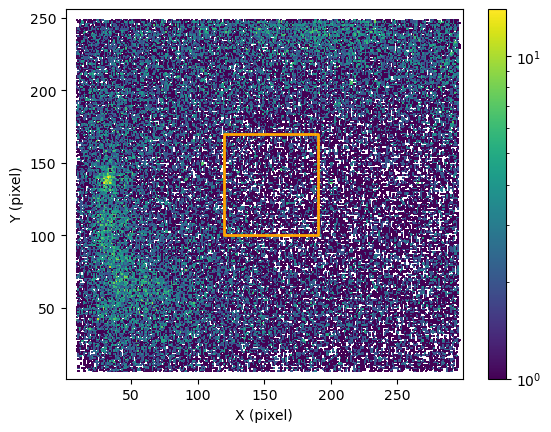}
    \caption{The sum of all $19$ runs taken with the magnetic field off and the cadmium absorber in place. The signal region of interest is outlined in the center of the plot.
}
    \label{fig:results:beam_0T_detect}
\end{figure}

To search for or place a limit on the transition regeneration probability for $n \rightarrow n' \rightarrow n$, a set of runs was taken with variable magnetic fields, with the Cd absorber in the center of the magnet to block the primary neutron beam, and the PC attenuators removed. 
The probability of oscillations into the mirror neutron state and back is enhanced when the energy splitting between the two states is minimized by the magnetic field compensating the mass-splitting.  
The regeneration effect should appear in the ROI of the detector as an excess of the neutron counting rate over the rate of background when magnetic field is present. The ROI was the same as used in the intensity determination. 
The summed data from the detector with the Cd absorber in place, formed by the combination of runs taken with no magnetic field at $0$~T, is shown in Figure~\ref{fig:results:beam_0T_detect}. 
The measured counting rate at $B=0$~T was used as a background to the measured signal at the $B = \pm 4.70$~T and $B = \pm 2.35$~T nominal magnetic field settings. 

The SNS runs at $60$~Hz, inducing a periodic background due to fast neutrons every $16666$~$\mu$s. 
To avoid this background, we placed a veto on events with a time of flight between $16661$~$\mu$s and $16711$~$\mu$s after the proton pulse signal. 
Data was collected and organized into runs with constant proton charge, such that the integrated charge on the accelerator target for a given run was $5.1$~C. 
Temporary outages of the accelerator were observed, resulting in longer runs and suppressed rates during the outages. 
Temporary outages longer than $10$~s affected $42$~\% of the typically 1 hour long runs.
The accelerator pulses were logged in the data and associated with proton pulse charge.
Every 10~s, an accelerator pulse is dropped for diagnostic purposes. 
During typical operation, the accelerator was stable such that each pulse deposited $23.32 \pm 0.07$~$\mu$C into the target. 
After a temporary outage, the accelerator took some time to ramp up, during which time the neutron intensity delivered from the moderator is expected to be nonlinear as a function of proton charge~\cite{Hugle:2018}. 
To mitigate this effect, accelerator pulses where the proton charge was less than $23$~$\mu$C were removed from the analysis. 
\begin{figure}
    \centering
    \includegraphics[width=\columnwidth]{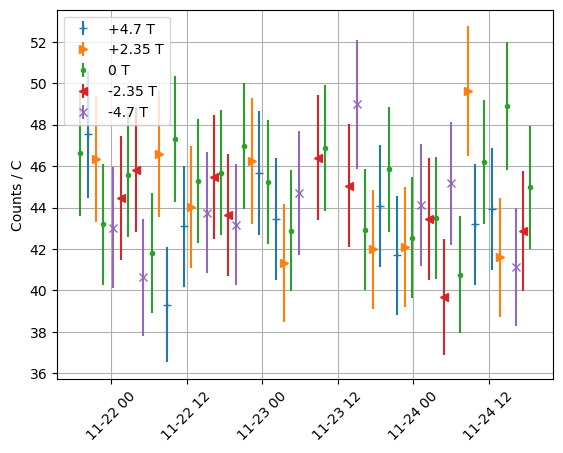}
    \caption{Measured neutron counts per Coulomb of proton beam charge inside the region on interest for each run. Runs are categorized by nominal magnetic field. (Color online)}
    \label{fig:results:measured_sig}
\end{figure}

After these cuts, the neutron counts measured in each individual run were normalized to the accumulated charge for that run. 
Normalized counts inside the region of interest for individual runs can be seen in Figure~\ref{fig:results:measured_sig}. 
Uncertainties on the raw counts inside the region of interest are assumed to follow a Poisson distribution, and the error bars indicate one standard deviation. 
The individual runs were combined to provide an overall signal for different magnetic fields, shown in Table \ref{tab:run_list_foreground}. 
\begin{table*}
    \centering
    \begin{tabular}{| c | c | c | c | c | c |}
        \hline
        $B$~Field ($T$) & Total Time (s) & Live Time (s) & ROI Counts & Total Charge (C) & Counts / C\\
        \hline
        $+ 4.70$ & $33147$ & $32559$ & $1987 \pm 46$ & $45.6$ & $43.5 \pm 1.0$ \\ 
        $+ 2.35$ & $32726$ & $32591$ & $2027 \pm 45$ & $45.6$ & $44.4 \pm 1.0$ \\ 
        $0.00$ & $81179$ & $68853$ & $4324 \pm 66$ & $96.3$ & $44.9 \pm 0.7$ \\ 
        $- 2.35$ & $32801$ & $32614$ & $2010 \pm 45$ & $45.6$ & $44.1 \pm 1.0$ \\ 
        $- 4.70$ & $39014$ & $32612$ & $1999 \pm 45$ & $45.6$ & $43.9 \pm 1.0$ \\ 
        \hline
    \end{tabular}
    \caption{Combined total signal in ROI region for each magnetic field nominal setting. Uncertainties are statistical and represent one statistical deviation.}
    \label{tab:run_list_foreground}
\end{table*}
From Figure \ref{fig:results:measured_sig} and Table \ref{tab:run_list_foreground}, one can see that background measurement with field $B=0$~T is consistent within error with the measurements with positive and negative values of magnetic field. 
This signifies that within measurement errors, no $n \rightarrow n' \rightarrow n$ is observed.
\subsection{Limits of \texorpdfstring{$\Delta{m}$}{Delta m} and \texorpdfstring{$\theta_{0}$}{theta} parameters of theory}
\label{ssec:limits}

Determination of the limit on $n \rightarrow n' \rightarrow n$ utilizes the Feldman-Cousins method for treatment of rare events~\cite{Feldman:1997qc}. 
The Feldman-Cousins method selects a likelihood range closest to the most likely one, and uses this to determine the $95$~\% confidence limit on the regeneration transition, $p_{tr.}$. 
These limits were calculated assuming Gaussian uncertainties for the signal and backgrounds. 
As mentioned above, background measurements come from the $0$~T data inside the ROI. 

\begin{table}[ht]
    \centering
    \begin{tabular}{|c|c|c|}
        \hline
        B-Field (T) & Signal ($\times 10^{-10}$) & $95$~\% C.L. $p_{tr.}$ ($\times 10^{-10}$) \\
        \hline
        $ 4.70$ & $-3.9 \pm 3.2$ & $3.1$\\
        $ 2.35$ & $-2.1 \pm 3.2$ & $4.4$\\
        $ 4.70$ and $2.35$ & $-3.1 \pm 2.7$ & $2.8$\\
        \hline
    \end{tabular}
    \caption{Limit on the $n \rightarrow n' \rightarrow n$ regeneration probability per neutron at 95$\%$ CL with Cd absorber for two measured magnetic field settings. The $\pm 4.70$~T and $\pm 2.35$~T datasets use both polarities with the $0$~T data taken as a background. The combined $4.70$ and $2.35$ dataset uses all constant field-on data, with the $0$~T data taken as a background.}
    \label{tab:results:limits}
\end{table}

A summary of the search for the exotic regeneration effect at different magnetic fields is presented in Table \ref{tab:run_list_foreground}. 
Since the neutron beam is not polarized, the positive and negative signs of the magnetic field can be averaged before the subtraction of the zero magnetic field background. 
The intensity of $(3.092 \pm 0.083)\times 10^9$~counts$/$C from Equation~\ref{eq:res:finalinten} is applied to determine the probability of the regeneration effect per initial neutron, averaged over the spectrum of neutron beam velocities. 
The corresponding Feldman-Cousins 95$\%$ confidence level limit is shown in Table \ref{tab:results:limits}. 
To relate this measured result of the excluded probability of transmission to the corresponding parameters $\Delta{m}$ and $\theta_{0}$ of the theoretical model \cite{Berezhiani:2018eds}, simulations of the neutron evolution through the experimental apparatus were performed. 

\subsection{Simulations of Neutrons in Matter and Magnetic Fields}
\label{ssec:simulations}

The probability of $n \rightarrow n'$ transmission through the absorber can be calculated for varying $\Delta{m}$ and $\theta_0$ parameters~\cite{Kamyshkov:2021kzi}. 
The density matrix Equation \ref{eq:phen:liouville_von_neumann}, containing the Hamiltonian defined in Equation \ref{eqn:phen:simp_hamiltonian}, was used to describe the evolution of the $(n,n')$ system through the 60\,cm path shown in Figure \ref{fig:exp:magneticField}, including the magnetic field and Cd absorber. 
For each initial neutron with given velocity $v$, $\Delta{m}$, and $\theta_0$, the evolution equation was solved numerically utilizing a custom GPU simulation package in the Julia programming language. 

The Liouville-von Neumann equation \ref{eq:phen:liouville_von_neumann} can be written using the matrix form $\dot{\rho} = \mathbf{M}\rho$, where within a sufficiently small step $dt = v dz$ the matrix $\mathbf{M}$ is constant. 
The solution to the differential equation can be found with the matrix exponential:
\begin{equation}
    \rho \left( t \right) = e^{\mathbf{M}t} \rho_{0}
    \label{eq:sim:mat_exp}
\end{equation}
The matrix exponential is calculated by finding the Jordan decomposition of the matrix $\mathbf{M}$:
\begin{equation}
    \mathbf{M} = S J S^{-1}
    \quad \quad
    e^{\mathbf{M} t} = S e^{J t} S^{-1}
    \label{eq:sim:mat_exp2}
\end{equation}

Since the density matrix $\hat{\rho}$ of the Liouville-von Neumann equation is Hermetian, equation \ref{eq:phen:liouville_von_neumann} can be rewritten as:
\begin{equation}
    \begin{pmatrix}
    \dot{\rho}_{11} \\ \dot{a} \\ \dot{b} \\ \dot{\rho}_{22}
    \end{pmatrix} = \frac{1}{\hbar} \begin{pmatrix}
    -2W & 0 & -2\epsilon & 0 \\
    0 & -W & +U & 0 \\
    \epsilon & -U & -W & -\epsilon \\
    0 & 0 & 2\epsilon & 0
    \end{pmatrix} \begin{pmatrix}
    \rho_{11} \\ a \\ b \\ \rho_{22}
    \end{pmatrix}.
    \label{eq:sim:tot_matrix}
\end{equation}
Equation \ref{eq:sim:tot_matrix} uses the Hermiticity of $\hat{\rho}$ to relate the off-diagonal components of the density matrix $\rho_{12} = a + ib$ and $\rho_{21} = a - ib$. 
Equation \ref{eq:sim:mat_exp2} can then be solved numerically for individual neutron trajectories with a fixed $\Delta{m}$, $\theta_0$, and velocity $v$ using the matrix described in Equation \ref{eq:sim:tot_matrix}. 
This calculation begins with a density matrix at $t=0$ in the pure neutron state:
\begin{equation}
    \label{eqn:sim:dens_mat}
    \hat{\rho}=\left(\begin{array}{cc}
        \rho_{11} & \rho_{12} \\
        \rho_{21} & \rho_{22} \\ 
    \end{array}\right)\
    =\left(\begin{array}{cc}
        1 & 0 \\
        0 & 0 \\ 
    \end{array}\right).\
\end{equation}

The potentials in Equation~\ref{eq:sim:tot_matrix} come from the presence of magnetic fields and material optical potentials. 
As the magnetic field $\vec{B}$ couples to the spin of the neutron $\vec{\mu}_n$, the calculation must be run for each spin $\pm \mu_{n}$ state separately and subsequently averaged.  
Equation~\ref{eqn:sim:dens_mat} defines the initial phase of the oscillating $(n,n')$ system at the time $t=0$ of the evolution calculation. 
The averaging over all initial phases is effectively provided by the averaging of the probability $(\rho_{11})$ of the neutron survival over the spectrum of neutron velocities, related to the time-of-flight spectrum in the Figure~\ref{fig:results:time_of_flight}. 

The simulation of the evolution through the variable magnetic field accounts for both neutron polarizations in the case of adiabatic transformation probability where $\Delta{m} > |\mu B_{max}|$ as well as the case of LZ transitions at the compensation point where $|\Delta{m}-\mu B|=0$. 
The magnetic field profile of Figure \ref{fig:exp:magneticField} was provided by the magnet manufacturer and converted into a cubic spline for interpolation. 
To avoid sharp steps in the magnetic field profile, this was extended out to a constant field of $50$~$\mu$T, accounting for the Earth's magnetic field. 
The mass splitting $\Delta{m}$ must be greater than $\mu B$ at the origin of the simulation in order for Equation \ref{eqn:sim:dens_mat} to be valid. 
The length of the neutron travel must thus be long enough to begin in a region of low field, below $10$~G. 
The rate of the magnetic field falloff outside the high field region does not significantly change the transmission probability. 

\begin{figure}
    \centering
    \includegraphics[width=\columnwidth]{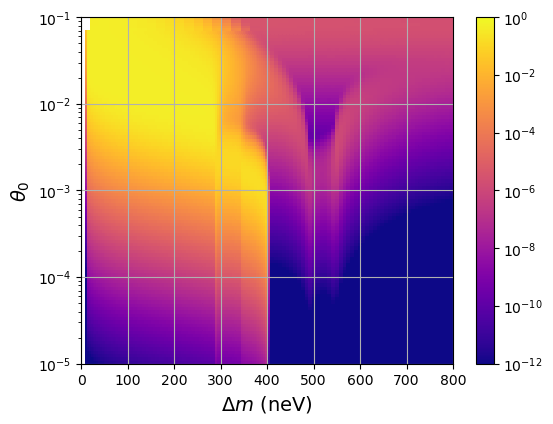}
    \caption{Simulated probability of regeneration (color scale) through $32$~mm of B$_4$C in a $4.8$~T field, the configuration used in \citep{Broussard:2021eyr}, as a function of model parameters $\Delta{m}$ and $\theta_0$. Here, $\Delta{m}$ is positive. Note the rapidly changing region between $400$ and $600$~neV due to the Fermi potential of B$_4$C.}
    \label{fig:sim:boroncarbide}
\end{figure}

\begin{table}[ht]
    \centering
    \begin{tabular}{|c|l|}
        \hline
        Material & ~~~~~~~~~~$V + i W$~(eV) \\
        \hline
        Cd & $5.88 \times 10^{-8} + i \cdot 8.46 \times 10^{-9}$\\  
        B$_4$C & $1.99 \times 10^{-7} + i \cdot 6.10 \times 10^{-9}$\\ 
        Air & $1.20 \times 10^{-10} + i \cdot  5.77 \times 10^{-15}$\\ 
        \hline
    \end{tabular}
    \caption{Material optical potentials used for calculating probability of neutron transmission through an absorber calculated from~\cite{nist_xsections} database.}
    \label{tab:sim:opt_pots}
\end{table}
Centered inside the magnetic field is a highly absorbing,  $3.5$~mm thick, Cd beamstop, with upstream surface aligned with the midpoint of the magnetic field. 
This differs from the previous experiment~\cite{Broussard:2021eyr} where a $32$~mm thick B$_4$C beamstop was used. 
The switch to Cd avoided a strong absorption resonance 
due to the large optical potential of B$_{4}$C (see \cite{Kamyshkov:2021kzi}, Table \ref{tab:sim:opt_pots} and an illustration in Figure \ref{fig:sim:boroncarbide}) which previously limited the sensitivity between 400 and 600 neV~\cite{Broussard:2021eyr}. 
In addition to the central absorber, the experimental apparatus was located  in air at room temperature, which has an additional optical potential. 
We simulated these materials with both real and imaginary parts of the material optical potentials in Equation \ref{eq:sim:tot_matrix} using values in Table~\ref{tab:sim:opt_pots}. 
The addition of air instead of vacuum outside of the absorber provides a $\sim 1$~$\%$ reduction in sensitivity for a given neutron trajectory. 

The velocity dependence was accounted for by averaging across $2000$ trajectories with fixed velocity, sampled from the velocity profile of neutrons coming from the SNS. 
This is numerically integrated with steps of constant $z = 30$~$\mu$m. 
Neutrons were simulated along a $2$~m total path, with the magnetic field centered in the simulation. 
The simulated probability for a given neutron trajectory downstream of the magnet at the end of the calculated evolution path has an oscillation behaviour similar to that of the $(n,n')$ system in vacuum. 
For this reason the probability of transmission was averaged over the final $3$~mm ($100$~steps). 
The step sizes and number of trajectories were varied to ensure numerical precision errors were avoided. 

\begin{figure*}
    \centering
    \includegraphics[width=\textwidth]{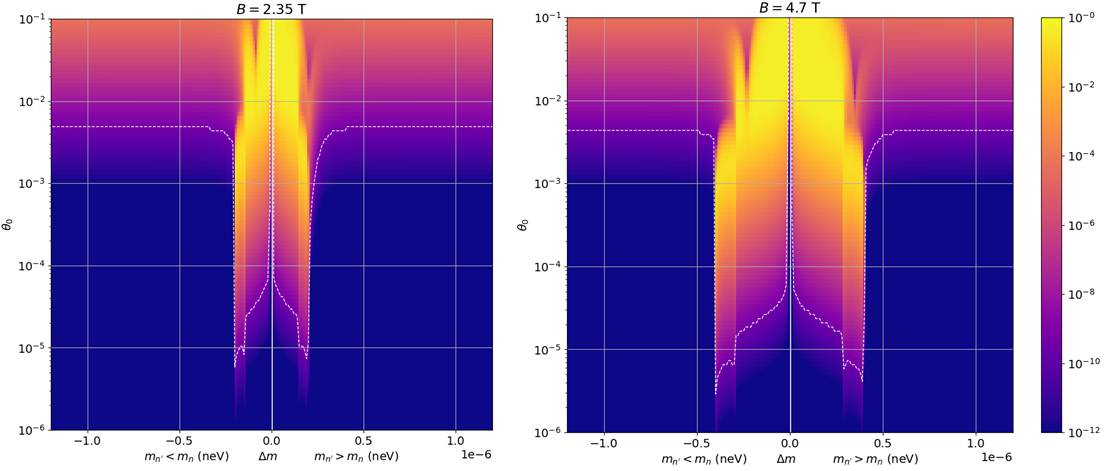}
    \caption{Simulated probability of regeneration (color scale)  through $3.5$~mm of Cadmium in a $2.35$~T (left) peak strength magnetic field and $4.7$~T (right) peak strength as a function of model \cite{Berezhiani:2018eds} parameters $\Delta{m}$ and $\theta_{0}$. Area above the dotted line is excluded at 95$\%$ CL by the sensitivity limit reported in Table \ref{tab:results:limits}.}
    \label{fig:sim:limits}
\end{figure*}

Systematic effects, such as positioning of the absorber inside of the magnetic field and the centering of the beam, were studied by running the simulation using an off-axis beam and by changing the location of the Cd absorber. 
Shifting the Cd absorber by $\pm 1$~mm inside the magnet did not change the simulated probability results shown in Figure \ref{fig:sim:limits}. 
Similarly, simulation of neutron evolution using the magnetic field $1$~cm off axis did not significantly affect the sensitivity in $\theta_0$. 

The results of the simulation described above for a Cd absorber with the two nominal magnetic fields, $4.70$~T and $2.35$~T, can be seen in figure \ref{fig:sim:limits}. 
The highest sensitivity to $n \rightarrow n' \rightarrow n$ regeneration can be found near the magnetic field peak. 
A reduced sensitivity can be seen for higher $\theta_0$ values for smaller $|\Delta{m}|$ and also when the $|\Delta{m}|$ exceeds the peak value of magnetic field. 
For values of $\Delta{m} > \mu_n \cdot B_{max}$, the sensitivity of the experiment is approximately  a constant value. 
For negative mass splittings, the sensitivity only changes by the exact position of the probability void position. 
At low mass splitting where $\vec{\mu}_{n} \cdot \vec{B} \gg \Delta{m}$, the frequency of oscillations becomes less dependent on the specifics of the field profile.

 \begin{figure}
    \centering   
    \includegraphics[width=\columnwidth]{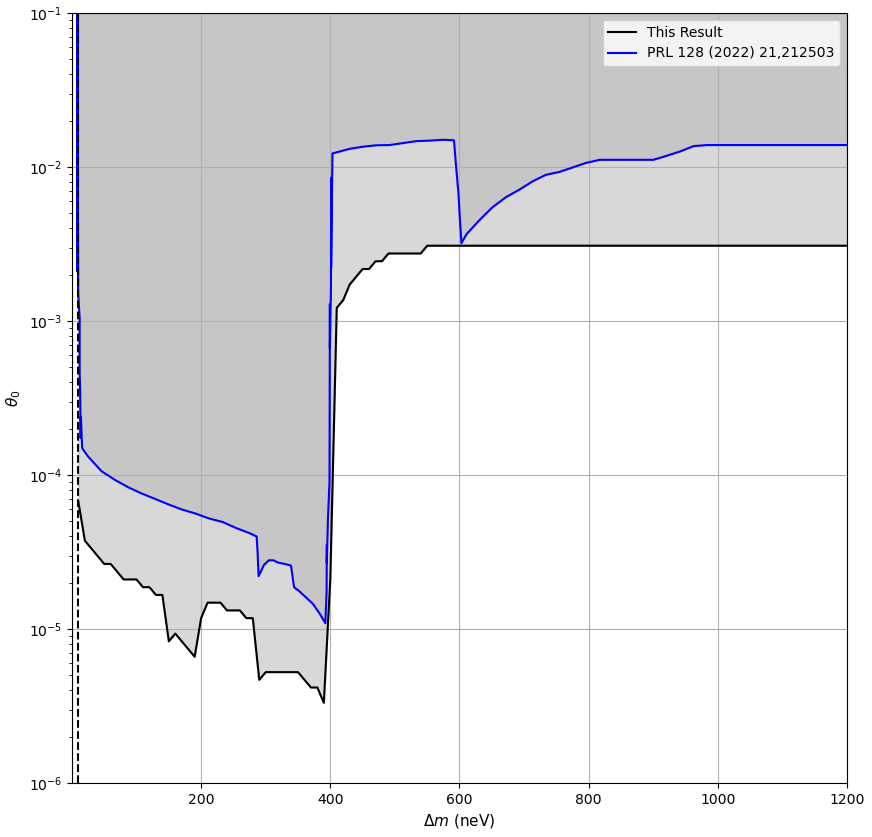}
    \caption{Comparison between the exclusion limits presented here to to the previous work~\cite{Broussard:2021eyr} for positive $\Delta{m}$. Note that in the previous result, the region between $400$~neV and $600$~neV was excluded due to the Fermi potential absorption resonance in B$_4$C.}
    \label{fig:results:old_vs_new}
\end{figure}

\begin{figure*}
    \centering
    \includegraphics[width=\textwidth]{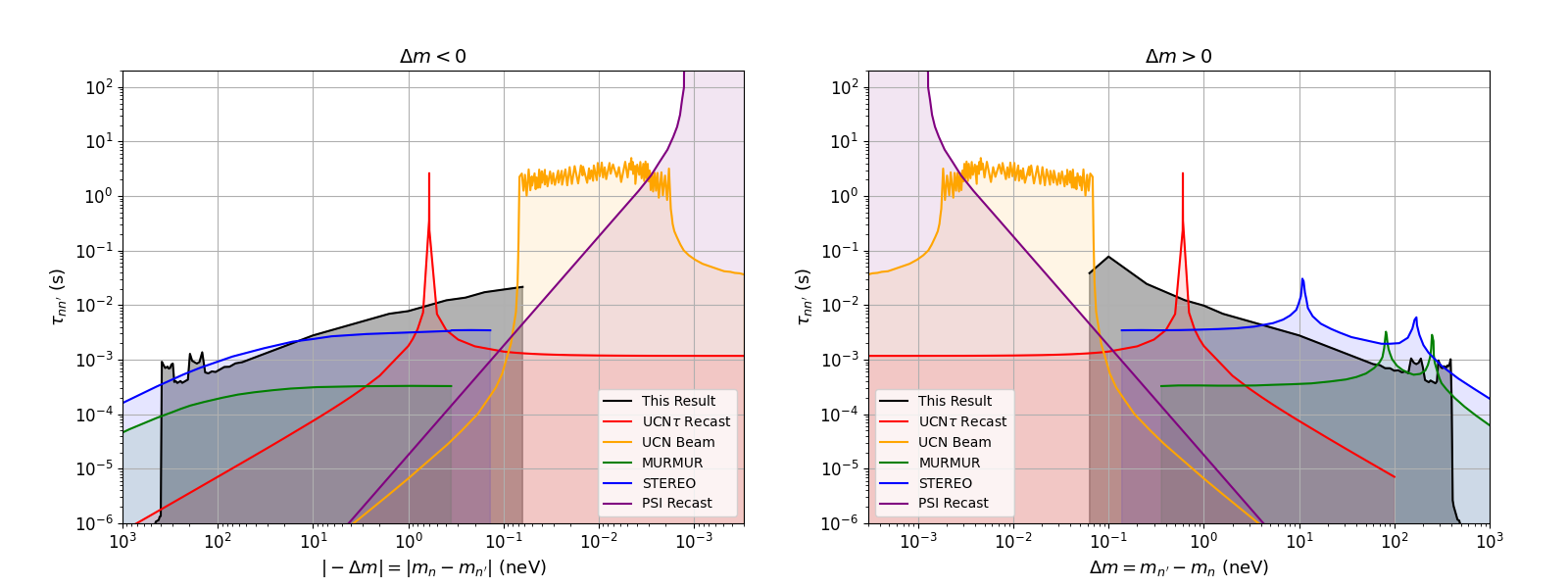}
    \caption{Limits reported in this paper, overlaid with other searches for sterile neutrons, assuming a unified framework as proposed in \cite{Hostert:2022ntu}. Results from UCN storage at PSI have been recast from searches using searches for a nonzero $\vec{B'}$~\cite{Ayres:2021zbh}. The limit from UCN$\tau$ comes from the non-observation of anomalous losses, as calculated in \cite{Hostert:2022ntu}. UCN beam results use disappearance in the GADGET detector~\cite{Ban:2023cja}. The STEREO and MURMUR reactor results are also presented~\cite{Stasser:2020jct,Almazan:2021fvo}.(color online)}
    \label{fig:results:actual_lims}
\end{figure*}

\begin{figure*}
    \centering
    \includegraphics[width=\textwidth]{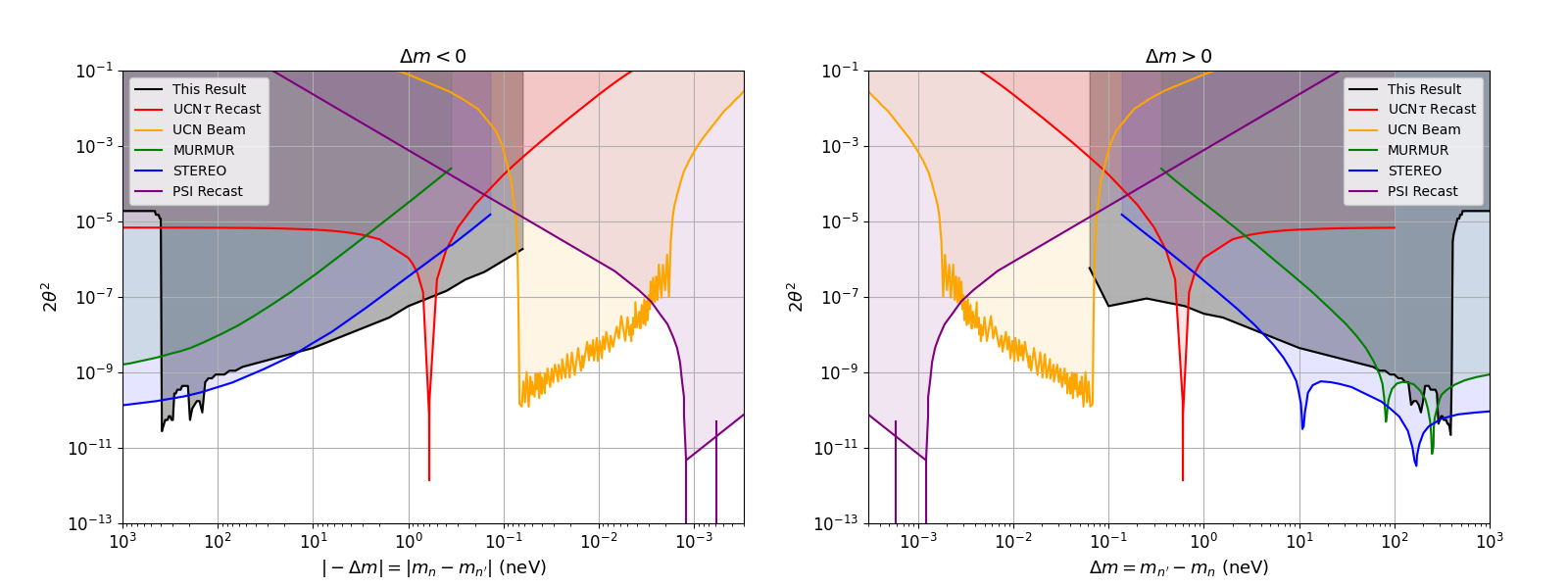}
    \caption{Same limits as in previous figure but in terms of $2\theta_0^2$, the probability of $n \rightarrow n'$ transformation in vacuum. (color online)} 
    \label{fig:results:compare_lims}
\end{figure*}

The overall parameter space excluded by this experiment can be found by taking the $95$~\% C.L. $p_{tr.}$ of Table~\ref{tab:results:limits} and superimposing this on the probability map of the $n \rightarrow n' \rightarrow n$ regeneration as function of parameters $\Delta{m}$ and $\theta_0$  for the respective magnetic fields. 
In calculating these limits, the regeneration effect was assumed to be fully responsible for any counting rate above background in the detector. 
The $\pm 4.7$~T and $\pm 2.35$~T results can be integrated with one another by finding the $95$~\% C.L. of exclusion of the combined probability map. 
This was done by determining a combined $\rho_{11}$ weighted by the sensitivity of the two results at every simulated point in $\Delta{m}$ and $\theta_0$:
\begin{equation}
    \rho_{11,tot.} = \frac{\rho_{(11,4.70)} / p_{(tr.;4.70)}^2 + \rho_{(11;2.35)} / p_{(tr.;2.35)}^2 }{1 / p_{(tr.;4.70)}^2 + 1 / p_{(tr.;2.35)}^2}
\end{equation} 
The Feldman-Cousins procedure can be used to find a new $95$~\% C.L. band for the combined signal from all magnet-on runs. 
The results can be seen in Figure~\ref{fig:results:old_vs_new}. 
The highest sensitivities are at the peak field strengths of the magnet and in the region of high magnetic field enhanced by the Landau-Zener effect. 
\section{Discussion}

\label{sec:conclusions}
This paper has reported the results of an experiment that provides the most stringent results found thus far for $n \rightarrow n'$ oscillations for the broad region $0.1$~neV $\leq \left| \Delta{m} \right| < 10$~neV and competitive results up to $400$~neV. 
Other experiments have used varied techniques in order to probe different mechanisms for $(n,n')$ transitions. 
These different experiments probe slightly different underlying sources of oscillation and use different analysis methodologies. 
The experiments can nevertheless be recast in terms of $\Delta{m}$ and compared to search across a wide parameter space. 
We follow a procedure synergistic with the framework for unifying these approaches proposed by~\cite{Hostert:2022ntu}. 
The results from this measurement are compared to other experimental searches in Figures~\ref{fig:results:actual_lims} and \ref{fig:results:compare_lims}. 
Since compensation for $\Delta{m}$ comes primarily from a magnetic field, the regeneration technique presented here is relatively insensitive to the sign of $\Delta{m}$. 
The technique presented here provides competitive limits on neutrons transitioning to mirror neutrons using existing neutron scattering instruments, requiring only a few days of dedicated beamtime.

The primary constraints of this experiment come from statistics, understanding of the neutron beam intensity, and understanding of the backgrounds.
This technique could be extended for use at higher intensity neutron sources, such as ORNL's High Flux Isotope Reactor (HFIR) or the European Spallation Source ~\cite{Abele:2022iml,Santoro:2023izd}. 
A subsequent search at a reactor source rather than a spallation source would potentially have a more stable neutron flux, providing improvements on the intensity determination. 
Studies done at higher flux would also require improved understanding of the backgrounds in the detector as well. 
Extending sensitivity to other energy splittings would require higher strength magnetic fields or better understanding of the low field range.

\section{Acknowledgements}
This research was sponsored by the U.S. Department of Energy (DOE), Office of Science, Office of Nuclear Physics [contract DE-AC05-00OR22725], by the Laboratory Directed Research and Development Program [project 8215] of Oak Ridge National Laboratory, managed by UT-Battelle, LLC, for the U.S. DOE, and in part by the U.S. DOE, Office of Science, Office of Work- force Development for Teachers and Scientists (WDTS) under the Science Undergraduate Laboratory Internship program. The research of the University of Tennessee, Knoxville group was partially supported by US DOE Grant DE-SC0023149. This research used resources at the Spallation Neutron Source, a DOE Office of Science User Facility operated by the Oak Ridge National Laboratory. The authors thank Michael Fitzsimmons for useful discussions.

\bibliography{references.bib}
\end{document}